\documentclass[10pt,a4paper,twocolumn]{article}
\usepackage[utf8]{inputenc}
\usepackage{amsmath, amssymb}
\usepackage{physics}
\usepackage[]{siunitx} %for easy & consistent units
	\DeclareSIUnit\molecule{molecule}
	\DeclareSIUnit\debye{D}
	\DeclareSIUnit\au{a.u.}
	\DeclareSIUnit\Buckingham{B}
\usepackage[top=40pt,bottom=40pt,left=48pt,right=46pt]{geometry}
\usepackage{graphicx}
\graphicspath{{./}}
\usepackage{afterpage}
\usepackage{amsmath}
\usepackage[super,sort&compress,comma]{natbib}

\usepackage{color}
\usepackage{authblk}

\newcommand{\exocross}{\textsc{ExoCross}}
\newcommand{\duo}{\textsc{Duo}}
\newcommand{\molpro}{\textsc{Molpro}}
\newcommand{\hitran}{\textsc{Hitran}}

\newcommand{\ai}{\textit{ab initio}}
\newcommand{\Ai}{\textit{Ab initio}}

\newcommand{\Xstate}{\ensuremath{X\,{}^{3}\Sigma^{-}}}
\newcommand{\astate}{\ensuremath{a\,{}^{1}\Delta}}
\newcommand{\bstate}{\ensuremath{b\,{}^{1}\Sigma^{+}}}
\newcommand{\cstate}{\ensuremath{c\,{}^{1}\Sigma^{-}}}
\newcommand{\dstate}{\ensuremath{d\,{}^{1}\Pi}}
\newcommand{\estate}{\ensuremath{e\,{}^{1}\Pi}}
\newcommand{\Astate}{\ensuremath{A\,{}^{3}\Pi}}
\newcommand{\Bstate}{\ensuremath{B\,{}^{3}\Sigma^{-}}}
\newcommand{\Cstate}{\ensuremath{C\,{}^{3}\Pi}}
\newcommand{\Cprimestate}{\ensuremath{C'\,{}^{3}\Pi}}
\newcommand{\Aprimestate}{\ensuremath{A'\,{}^{3}\Delta}}
\newcommand{\Aprimeprimestate}{\ensuremath{A''\,{}^{3}\Sigma^{+}}}

\newcommand{\tsub}[1]{\textsubscript{#1}} %shorter subscript
\newcommand{\wav}[1]{#1 cm$^{-1}$}
\newcommand{\cm}{cm$^{-1}$}

\newcommand{\band}[2]{$#1\xrightarrow[]{}#2$}
\newcommand{\brkt}[3]{$\bra{#1}#2\ket{#3}$}
\newcommand{\brkteq}[3]{\bra{#1}#2\ket{#3}}

\usepackage{blindtext}
\usepackage{multicol}

\title{An ab initio study of the rovibronic spectrum of sulphur
monoxide (SO): diabatic vs. adiabatic representation}
\author[1]{R.P. Brady}
\author[1]{S.N. Yurchenko}
\author[2]{G.-S. Kim}
\author[1]{W. Somogyi}
\author[1]{ J. Tennyson}

\affil[1]{\textit{Department of Physics and Astronomy, University College London, Gower Street, WC1E 6BT London, United Kingdom}}
\affil[2]{\textit{Dharma College, Dongguk University, 30, Pildong-ro 1-gil, Jung-gu, Seoul 04620, Korea}}

\date{\today}

\begin{document}

\twocolumn[
  \begin{@twocolumnfalse}
    \maketitle
    \begin{abstract}
        We present an \textit{ab initio} study of the rovibronic spectra of sulphur monoxide ($^{32}$S$^{16}$O) using internally contracted multireference configuration interaction (ic-MRCI) method and aug-cc-pV5Z basis sets. It covers 13 electronic states \Xstate, \astate, \bstate, \cstate, \Aprimeprimestate, \Aprimestate, \Astate, \Bstate, \Cstate, \dstate, \estate, \Cprimestate, and $(3)^{1}\Pi$ ranging up to \wav{66800}. The \ai\ spectroscopic model includes 13 potential energy curves, 23 dipole and transition dipole moment curves, 23 spin–orbit curves, and 14 electronic angular momentum curves. A diabatic representation is built by removing the avoided crossings between the spatially degenerate pairs \Cstate -- \Cprimestate\ and \estate -- $(3)^1\Pi$ through a property-based diabatisation method. We also present non-adiabatic couplings and diabatic couplings for these avoided crossing systems. All phases for our coupling curves are defined, and consistent, providing the first  fully reproducible spectroscopic model of SO covering the wavelength range longer than 147 nm. Finally, an \textit{ab initio} rovibronic spectrum of SO is computed.
        
    \end{abstract}
  \end{@twocolumnfalse}
]

\section{Introduction}
\label{sec:Introduction}
Sulphur monoxide (SO) has many experimental and observational applications, making it a molecule of continuing interest in spectroscopic and kinetic studies. SO is an important intermediate in the oxidation of sulphur compounds within combustion reactions, \citep{48Gaydon.book} making SO studies applicable to environmental chemistry since sulphur oxides lead to the acidification/pollution of Earth’s atmosphere.\citep{04Wang,17Thurston,19Pan,11PAN} \citet{87BuLoHa.SO} show the importance  of SO within chemical reactions of Earth’s atmosphere due to its high reactivity and chemical involvement with N\tsub{2} and O\tsub{2}. Additionally, UV lasing in SO was demonstrated through optically pumping transitions within the \Bstate\ -- \Xstate electronic band \citep{91MiYaSm.SO} and then later for UV energy storage lasers. \citep{92StCaPo.SO} SO also has astrophysical importance, its Zeeman splitting has been used to probe the magnetic fields of dense star-forming regions, such as the Orion molecular cloud, for field strengths $\geq 10^{-2}$ G \citep{74ClJoxx.SO,17CaLaCo.SO} and its presence within warm chemistry's mean it is an excellent shock tracer. \citep{93ChMaxx.SO, 91AmElEl.SO}

% FOR LINE LIST PAPER: sulphur monoxide (SO) has been observed in many astronomical environments, including the interstellar medium, molecular clouds, star forming regions, and planetary atmospheres \citep{78GoGoLi.SO, 97CoMuxx.SO, 96Lellouch, 02deRoGr.SO,90NaEsSk.SO, 12BeMoBe.SO,87BlSuMa.ISM, 78GoGoLi.SO, 87BlSuMa.ISM}. Numerous studies propose sulphur-bearing molecules, including SO, as constituents of volcanic planetary atmospheres \citep{98ZoFexx.SO,21HoRiSh, 12Krasnopolsky}. It also plays a role in many solar-system atmospheres, including that of Jupiter’s moon Io \citep{96Lellouch, 02deRoGr.SO} and of Venus, \citep{90NaEsSk.SO, 12BeMoBe.SO}, as well as Earth's own atmosphere \citep{87BuLoHa.SO}. SO also has important experimental applications in spectroscopy, such as UV lasing \citep{91MiYaSm.SO, 92StCaPo.SO}.

The frontier orbitals of SO resemble that of carbon monoxide where the two $\pi^{*}$ valence electrons mean SO has a \Xstate\ ground state similar to O\tsub{2} and S\tsub{2}. Being isoelectronic with O\tsub{2}, SO has two low-lying metastable \astate\ and \bstate\ states which are relatively short lived due to large spin-orbit coupling. SO has been subject to pure rotational, \citep{94KlBeSa.SO,93Yamamoto.SO, 15MaHiMo.SO} electronic, \citep{68Colin.SO,82Colin.SO,99SeFiRa.SO,86ClTexx.SO,94StCaPo.SO} and ro-vibrational \citep{85KaBuKa.SO,82WoAmBe.SO,88KaTiHi.SO} experimental spectroscopic studies. More recently, \citet{21BeJoLi.SO,22BeJoLi.SO} produced  empirical line lists for the  $b\,{}^{1}\Sigma^{+}$--$X\,{}^{3}\Sigma^{-}$ and $a\,{}^{1}\Delta$--$X\,{}^{3}\Sigma^{-}$ rovibronic bands and  for the $X\,{}^{3}\Sigma^{-}$ and $a\,{}^{1}\Delta$ rovibrational bands of SO. The literature also contains many theoretical studies of SO, the most comprehensive, and one we compare to often, being \citet{11YuBixx.SO} who give spectroscopic constants on all electronic states we consider here (except $(3)^1\Pi$) computed through internally contracted multireference configuration interaction (ic-MRCI) methods using aug-cc-pV5Z basis sets. Another important theoretical work is by \citet{19SaNaxx.SO} who study the UV region of SO non-adiabatically where they compute PECs, DMCs, and non-adiabatic couplings (NACs) for the \Xstate, \Astate, \Bstate, \Cstate, \Cprimestate, $(3)^3\Sigma^-$, $(4)^3\Pi$, and $(5)^3\Pi$ states at a MRCI-F12+Q level of theory using aug-cc-pV(5+d)Z basis sets. \citet{19SaNaxx.SO} are also the first to compute cross-sections for SO longer than 190~nm in the UV.

% LINE LIST PAPER: For most of these spectra, SO was recorded in non-local thermodynamic equilibrium (non-LTE) conditions, and so we compare only relative intensities at best. 

Despite the strong experimental, theoretical, and observational baseline for SO, to this date the existing spectroscopic line list data for SO are limited in coverage. HITRAN \citep{HITRAN2012} contains only transitions between electronic states \Xstate, \astate\ and \bstate considering relatively low vibrational excitations. SO data and spectroscopic databases NIST \citep{NIST} and CDMS \citep{CDMS} contain data covering the microwave region only. In this study we aim to provide a strong theoretical base from which an accurate spectroscopic description of SO with large coverage can be made through the refinement of our \ai\ model to empirically determined energies in a future work. To achieve this we calculate \ai\ potential energy curves (PECs), spin-orbit curves (SOCs), electronic angular momentum curves (EAMCs), and electric (transition) dipole moment curves ((T)DMCs) for the 13 lowest electronic states of SO (\Xstate, \astate, \bstate, \cstate, \Aprimestate, \Aprimeprimestate, \Astate, \Bstate, \Cstate, \dstate, \estate, \Cprimestate, $(3)^1\Pi$) at an MRCI level of theory using aug-cc-pV5Z basis sets. The relative phases  of the  the non-diagonal couplings and transition dipole moments provided are fully self-consistent, \citep{14PaHiTe.AlO} which is crucial for  reproducible spectroscopic studies. 

Our \ai\ curves of SO are adiabatic as computed under the Born-Oppenheimer approximation\citep{27BoOpxx.diabat} and so the spatially degenerate states \estate, $(3)^1\Pi$ and \Cstate, \Cprimestate\ of SO exhibit avoided crossings due to their shared symmetries, where the non-adiabatic effects play important role. The associated non-adiabatic couplings (NACs) arise from the nuclear motion kinetic energy operator acting on electronic wavefunctions and contain first- and second-order differential operators with respect to the nuclear coordinates. Inclusion of NACs within numerical dynamics calculations are computationally costly when using analytical forms for the PECs and couplings because of both the cusp-like nature of PECs and the NACs having singularities at the region of spatial degeneracy. \citep{82MeTrxx.diabat,04JaKeMe.diabat, 18KaBeVa.diabat,21ShVaZo.diabat} This makes the fitting of analytical forms to the PECs and couplings almost impossible. Here we explore a diabatisation procedure to transform the adiabatic basis to the diabatic basis using NACs reconstructed from PECs as opposed to obtaining them \textit{ab initio} \citep{19SaNaxx.SO}, where the adiabatic first-order differential couplings are negligible or vanish, at the cost of inducing off-diagonal potential energy couplings, \citep{81Delos.diabat} or diabatic couplings (DCs).\citep{82MeTrxx.diabat, 04JaKeMe.diabat} In this representation, the electronic wavefunctions are smooth which lessens both numerical problems encountered in calculations within the adiabatic representation and the computational cost in computing NACs. Diabatisation helps recover non-Born-Oppenheimer dynamics \citep{04JaKeMe.diabat,03NaTrxx.diabat} and is important for processes such as collisions between open-shell molecules with spatially degenerate electronic states.\citep{18KaBeVa.diabat, 16KaVaGr.diabat, 14QiJaMi.diabat, 17JaQiMi.diabat, 17DeKaVo.diabat} The importance of the non-adiabatic effects on the spectral properties of SO is analysed. 

\section{Computational Details}
\label{sec:Computational Details}

%\subsection{\Ai\ Calculations}
%\label{sec:ai_calculations}

\begin{figure}[!]
    \centering
    \includegraphics[width=0.48\linewidth]{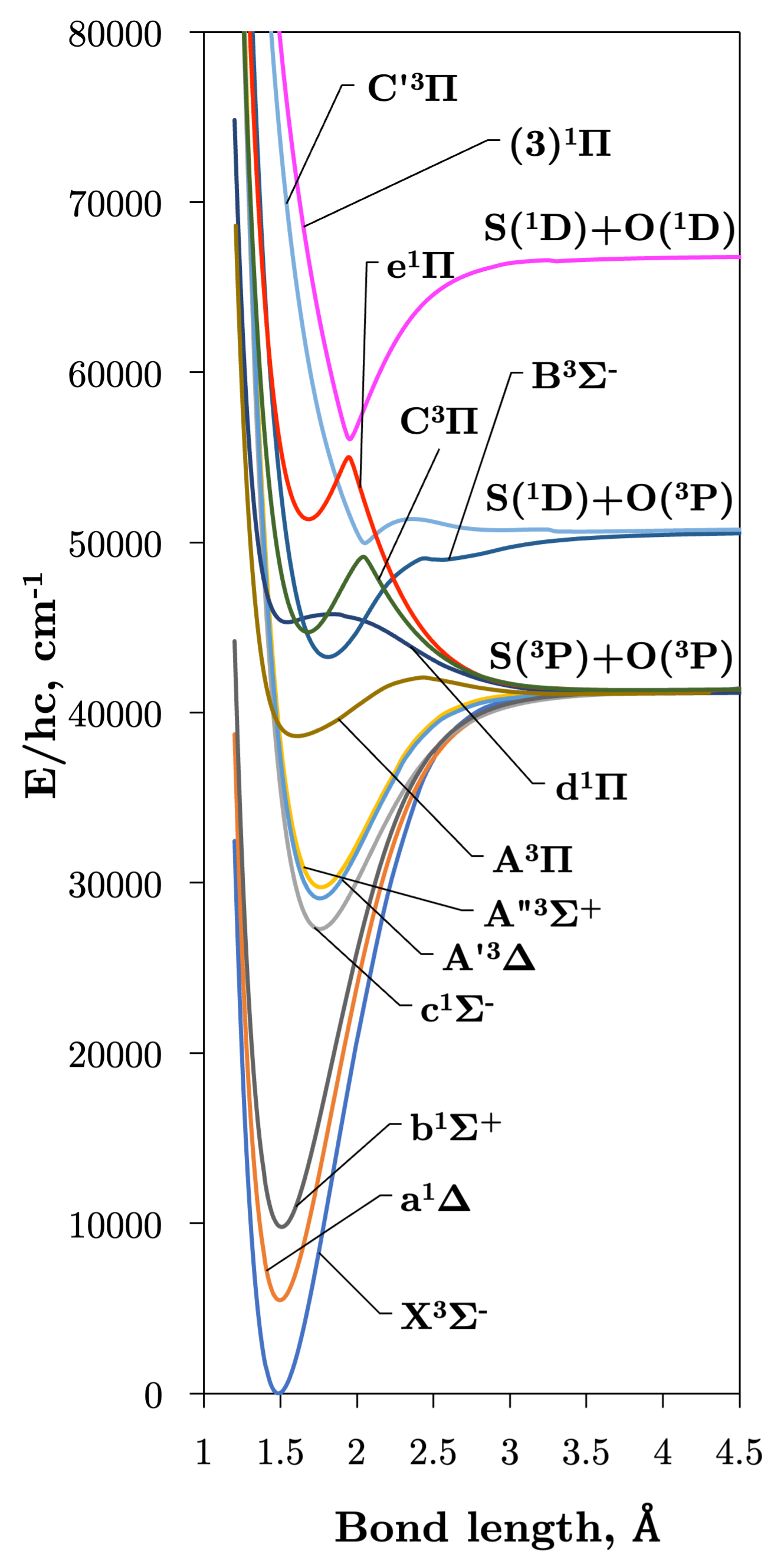}
    \includegraphics[width=0.48\linewidth]{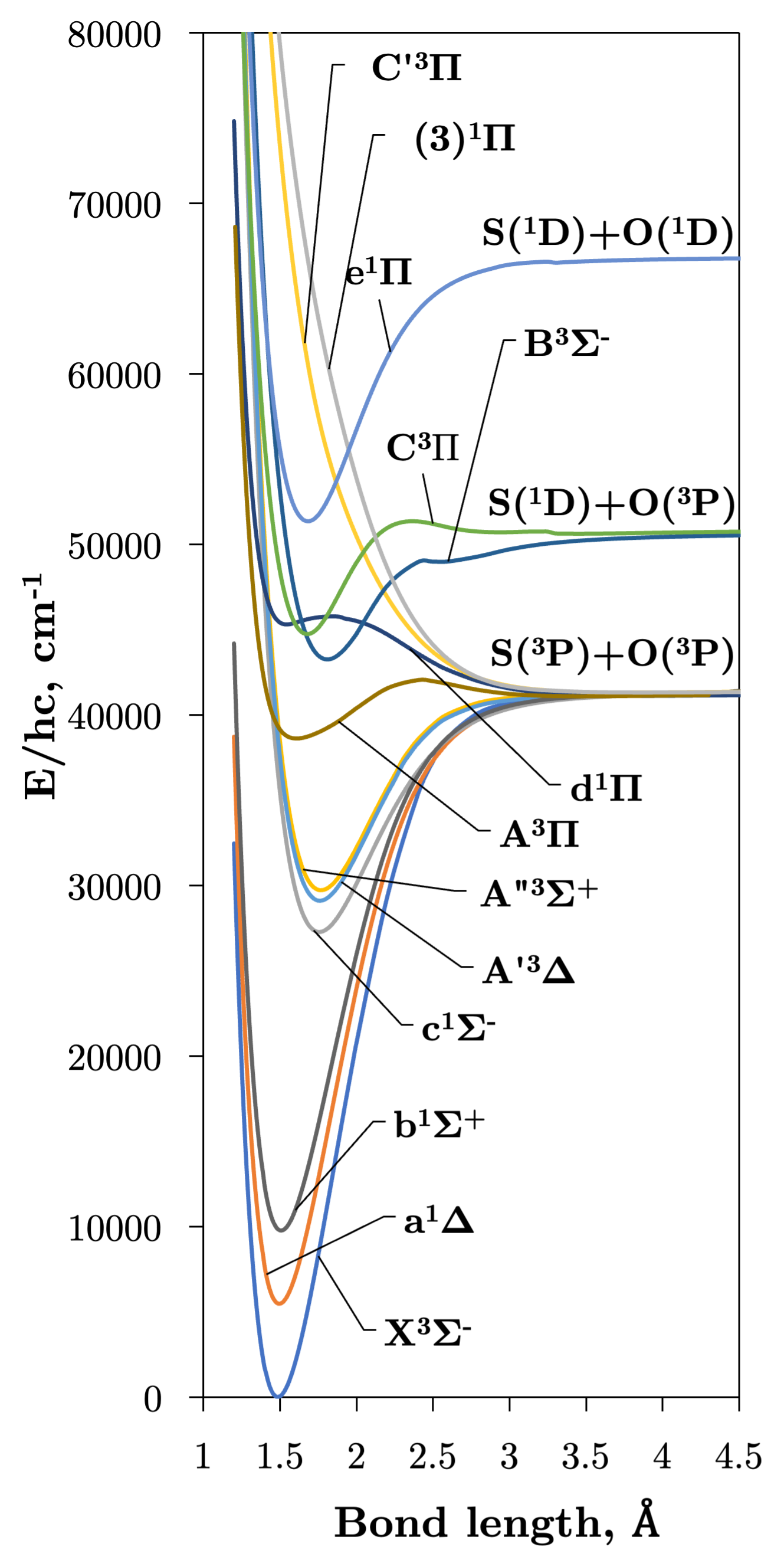}
    \caption{Plots of our computed 13 \ai\ PECs covering the first 7 triplet and 6 singlet electronic states up to $80,000$ \cm\ over inter-nuclear separations 1.0 \AA\ — 4.5 \AA. The adiabatic and diabatic representations of the PECs are shown in the left and right hand panels respectively.}
    \label{fig:ai_PECs}
\end{figure}

\begin{figure}[h!]
    \centering
    \includegraphics[width=0.48\linewidth]{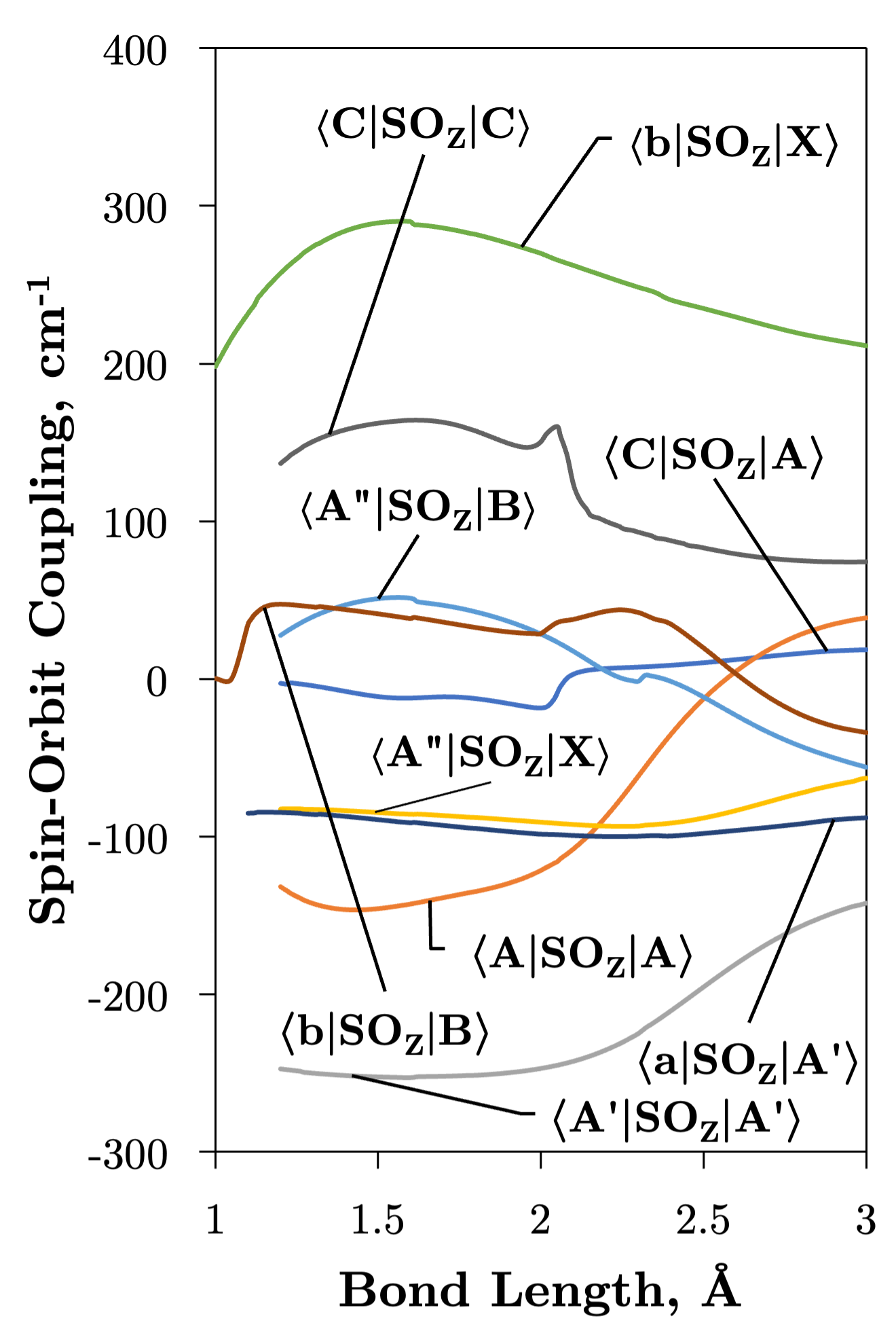}
    \includegraphics[width=0.48\linewidth]{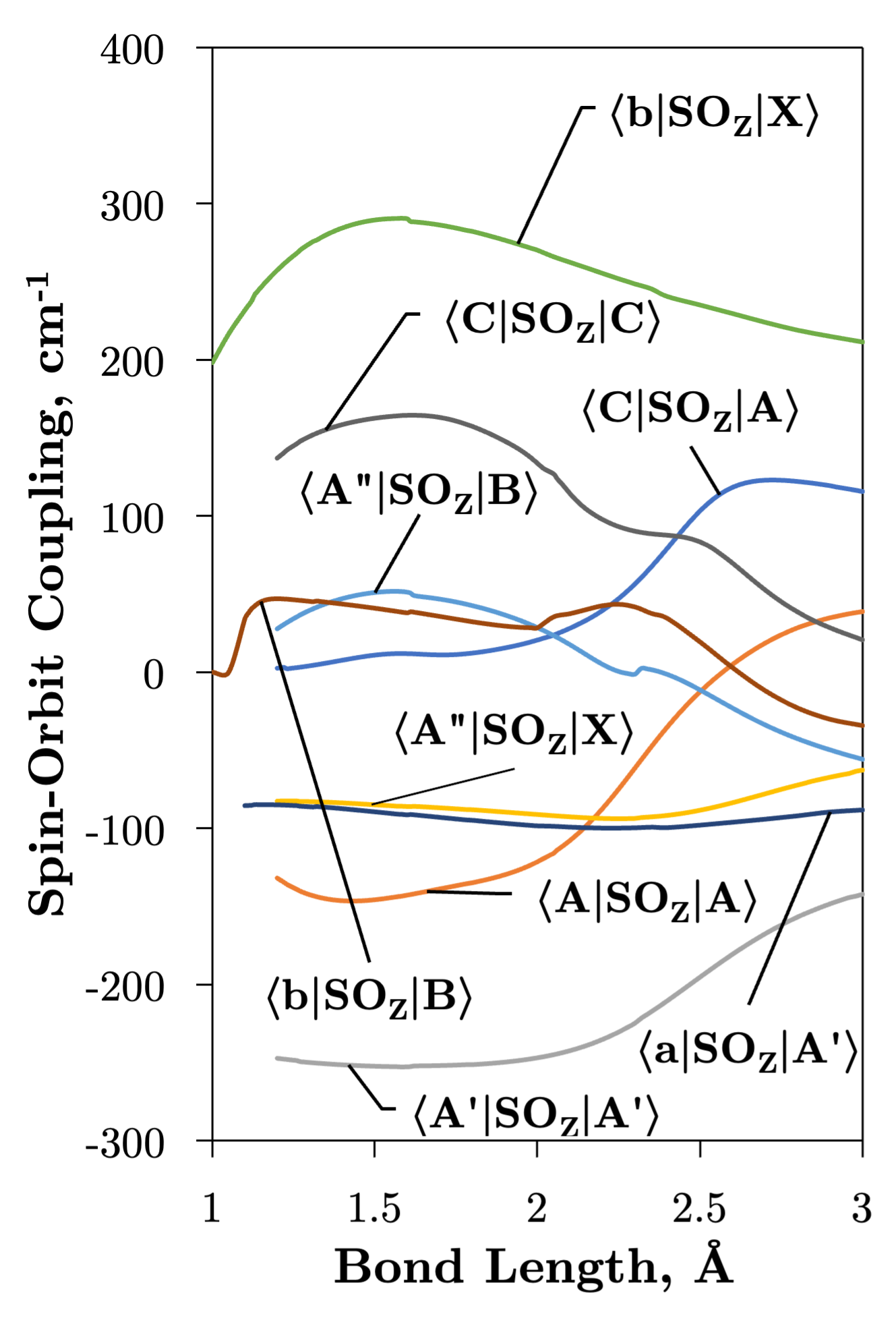}
    \caption{\Ai\ spin-orbit curves between states of the same $\Lambda$ (projection of the angular momentum) as a function of bond length. 
    In general, the spin-orbit coupling is strong also between states of different multiplicity. 
    These curves have been multiplied by $-i$ for visual purposes.}
    \label{fig:ai_LSZ}
\end{figure}

\begin{figure}[h!]
    \centering
    \includegraphics[width=0.48\linewidth]{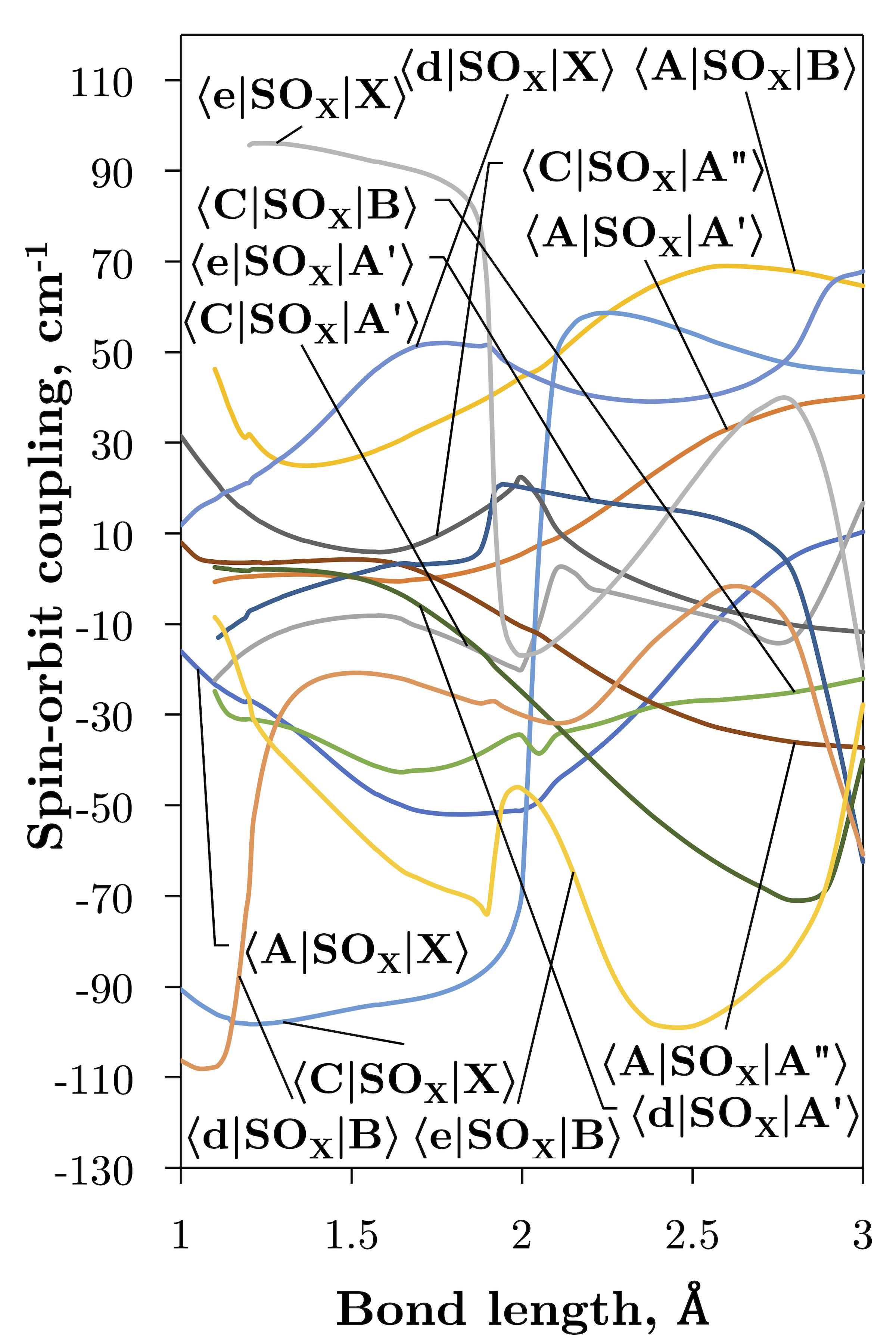}
    \includegraphics[width=0.48\linewidth]{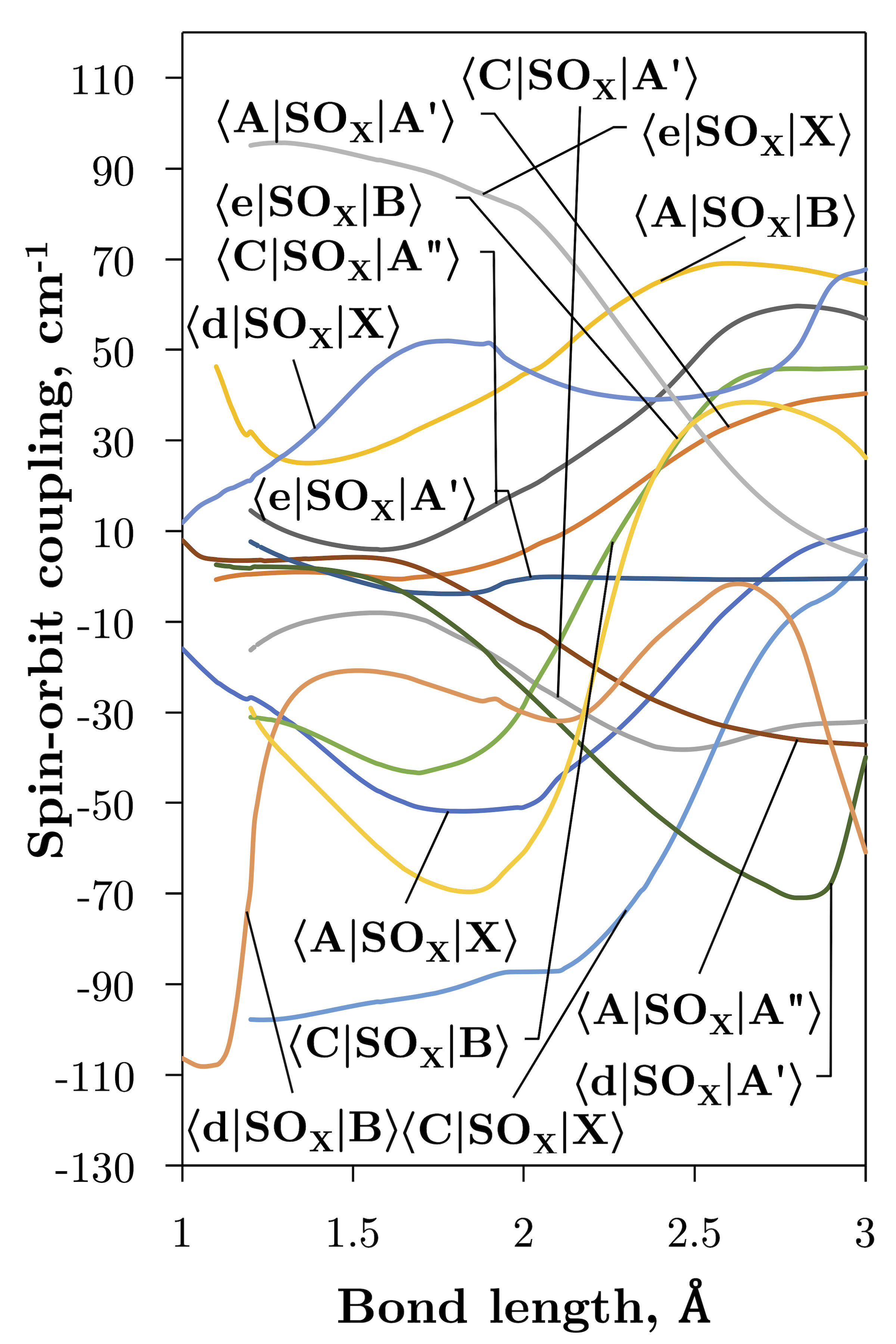}
    \caption{\Ai\ spin-orbit MOLRPO matrix elements in the adiabatic (left) and diabatic (right) representations between states of different values of $\Lambda$ as a function of bond length. These curves have been multiplied by $-i$ for visual purposes.}
    \label{fig:ai_LSX}
\end{figure}

\begin{figure}[h!]
    \centering
    \includegraphics[width=0.48\linewidth]{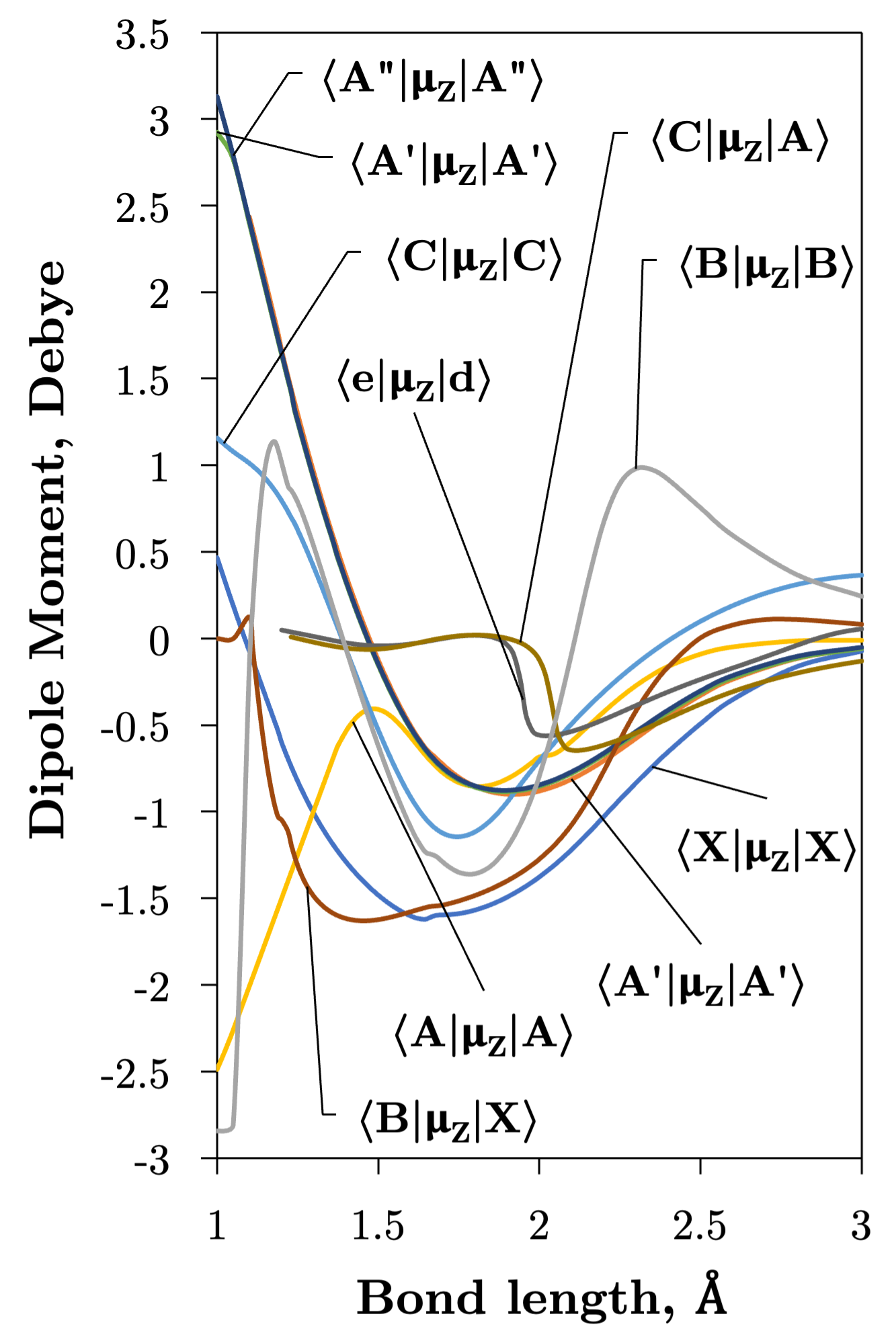}
    \includegraphics[width=0.48\linewidth]{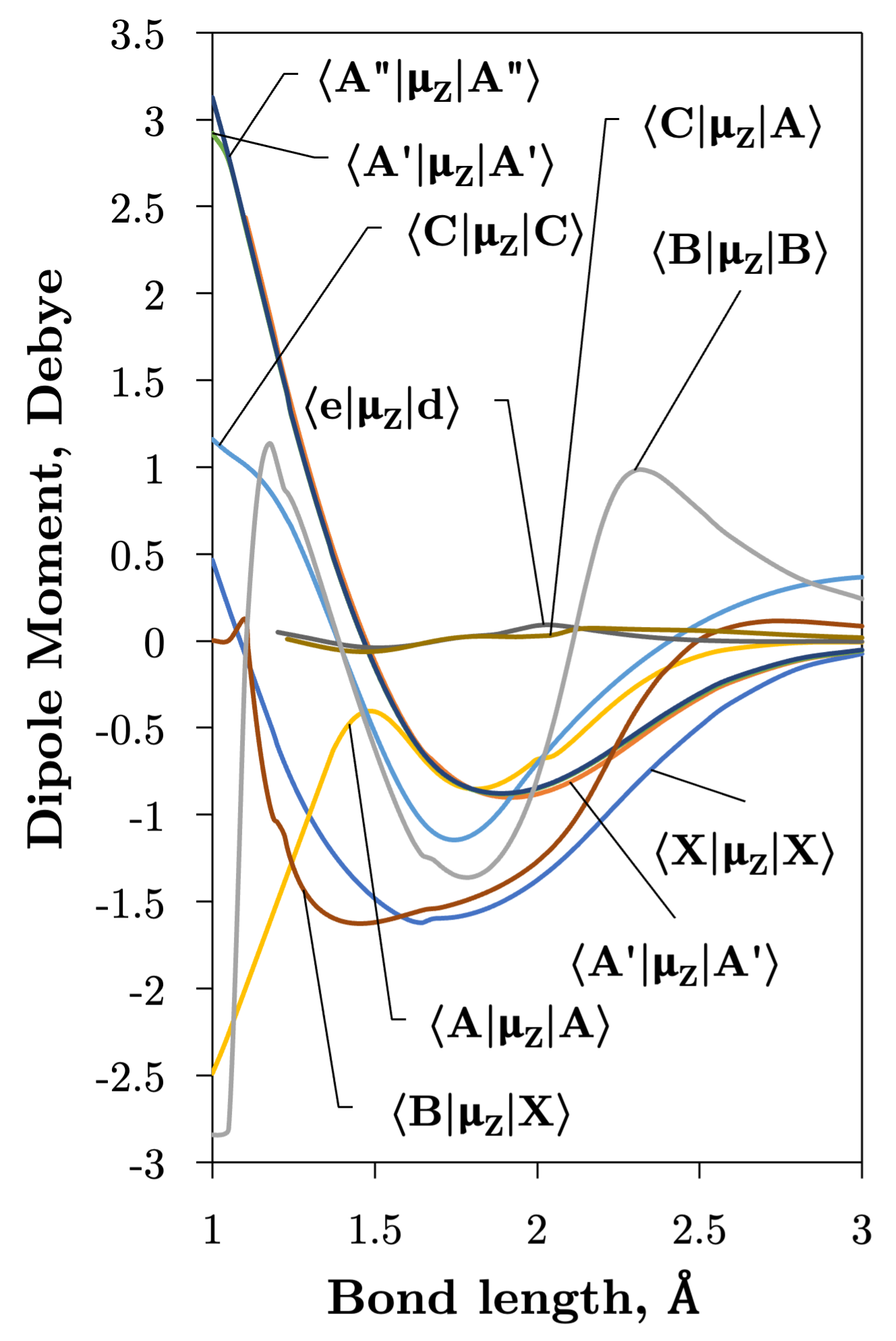}
    \caption{\Ai\ (transition) dipole moment matrix elements (Debye) between states of the same symmetry ($\Lambda$ and multiplicity) in the adiabatic (left) and diabatic (right) representations as a function of the bond length (\AA).}
    \label{fig:ai_DMZ}
\end{figure}

\begin{figure}[h!]
    \centering
    \includegraphics[width=0.48\linewidth]{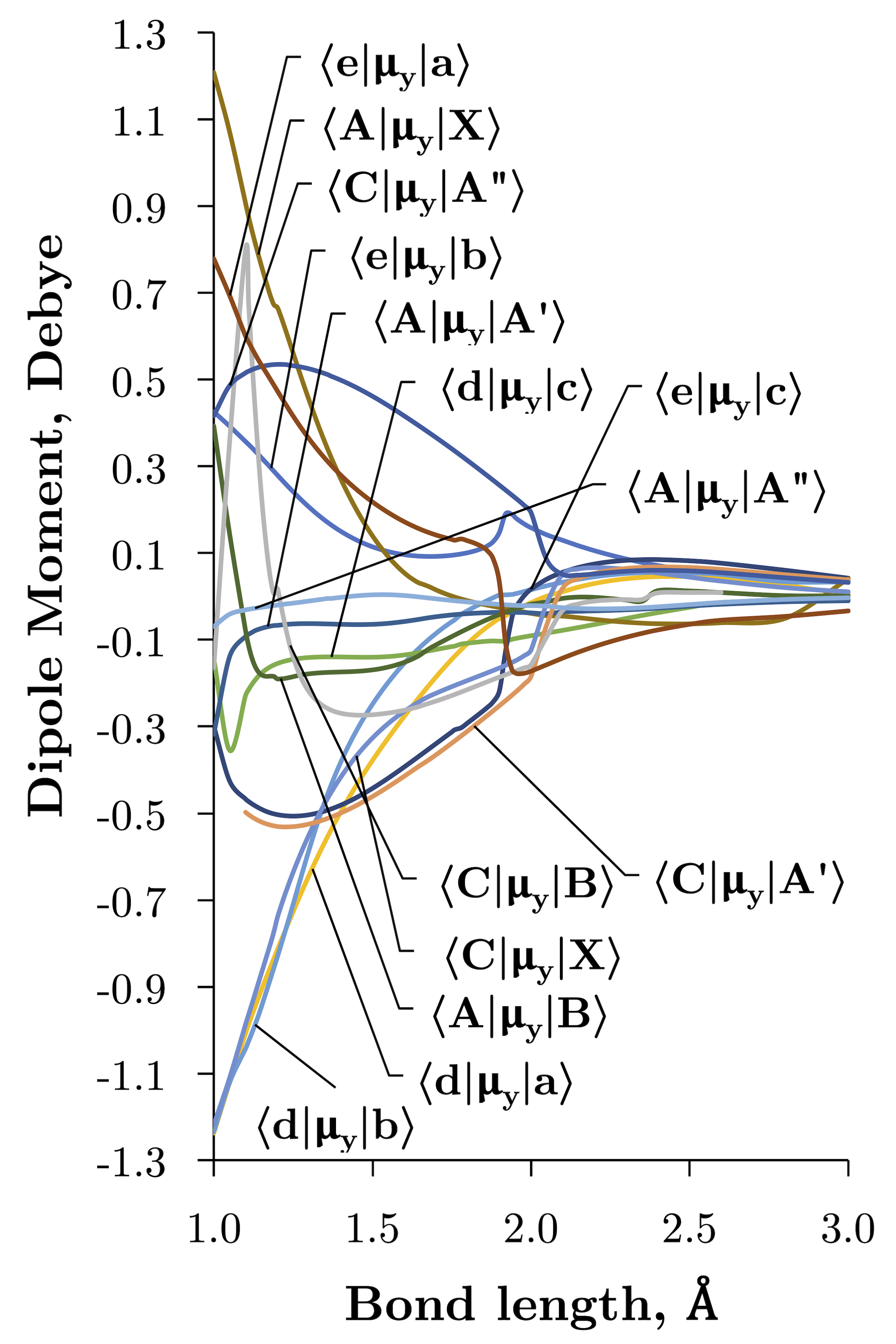}
    \includegraphics[width=0.48\linewidth]{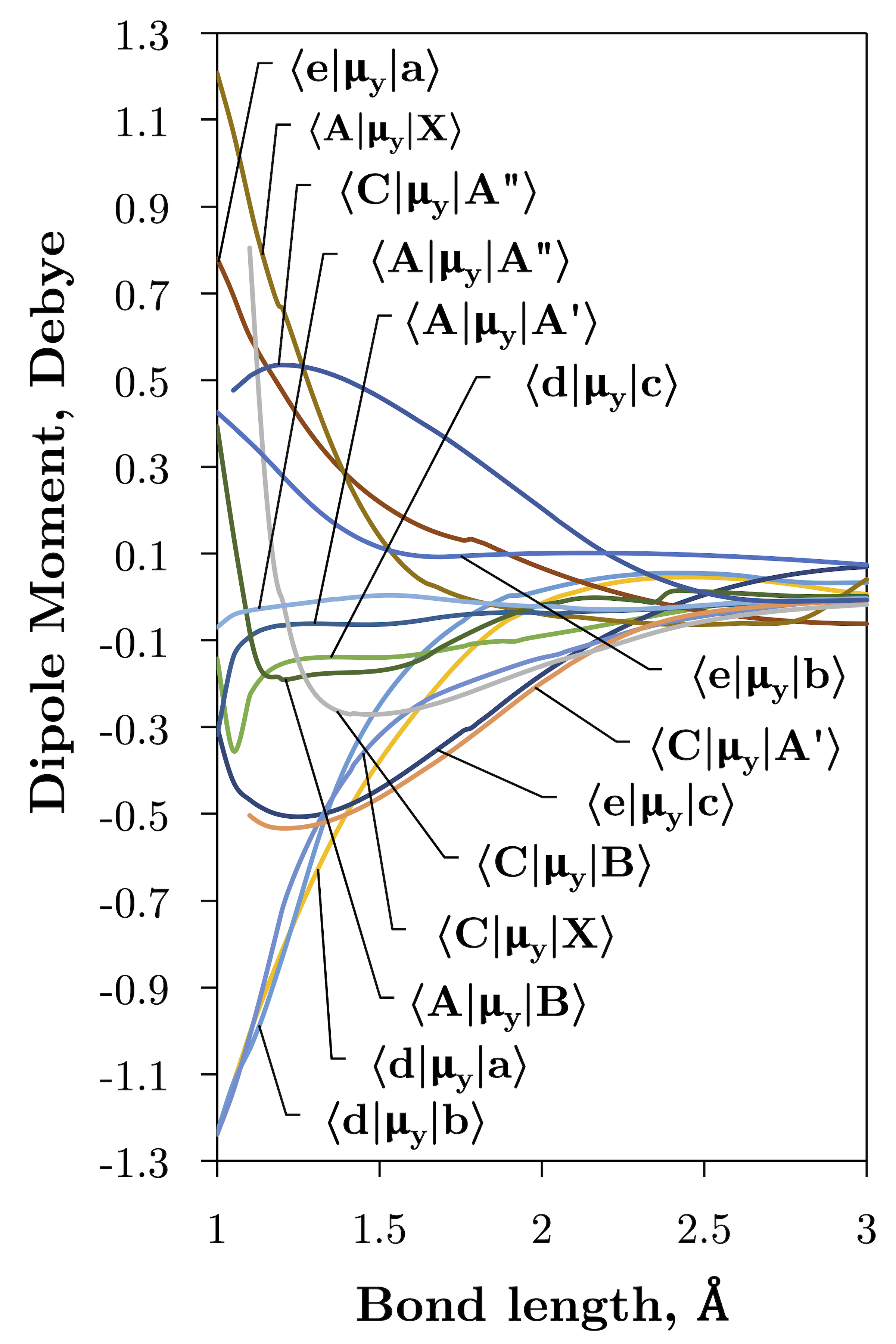}
    \caption{\Ai\ transition dipole moment curves  (Debye) between states of different symmetry in the adiabatic (left) and diabatic (right) representations as a function of the bond length (\AA).}
    \label{fig:ai_DMY}
\end{figure}

\begin{figure}[h!]
    \centering
    \includegraphics[width=0.48\linewidth]{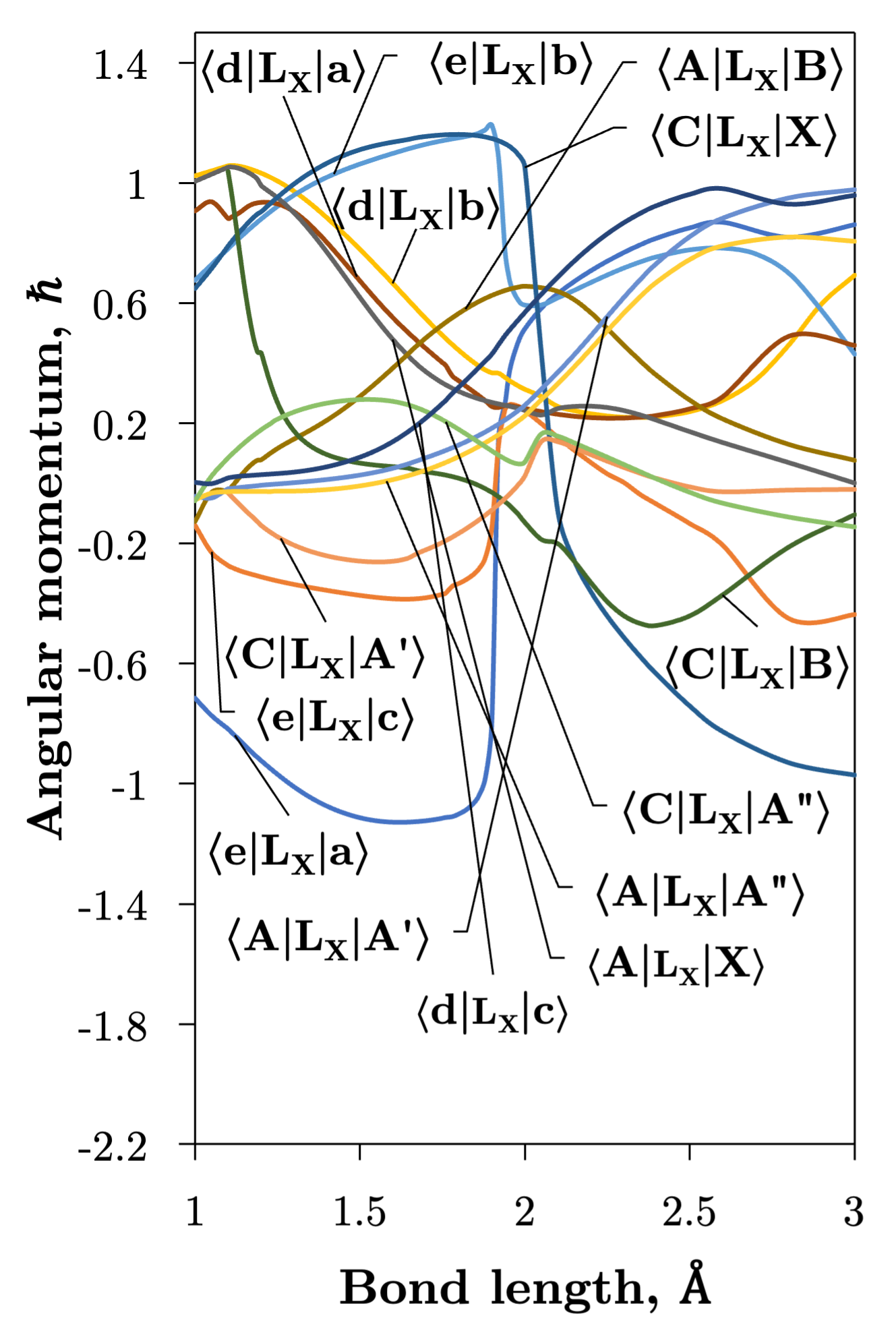}
    \includegraphics[width=0.48\linewidth]{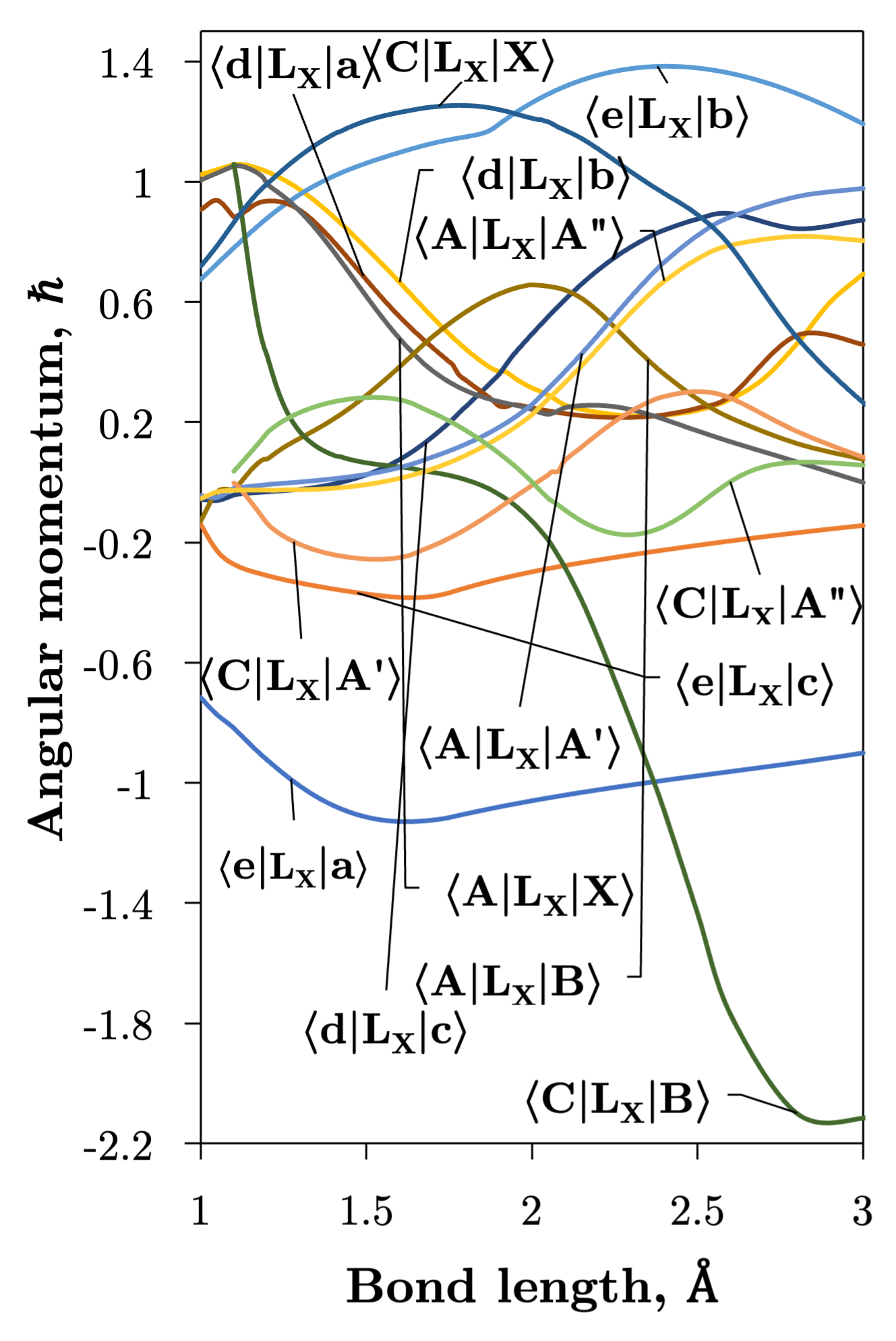}
    \caption{\Ai\ electronic angular momentum curves  in the adiabatic (left) and diabatic (right) representation plotted over bond length. $L_x$ means the $x$-component of the electronic angular momentum. These curves have been multiplied by $(-i)$ for visual purposes.}
    \label{fig:ai_EAMCs}
\end{figure}

Internally-contracted multi-reference configuration interaction (icMRCI) \ai\ calculations for the 13 lowest states of SO correlating with S(\textsuperscript{3}P) + O(\textsuperscript{3}P), S(\textsuperscript{1}D) + O(\textsuperscript{3}P) and S(\textsuperscript{1}D) + O(\textsuperscript{1}D) were performed using \molpro\ \citep{MOLPRO2015} with aug-cc-pV5Z basis sets\citep{89Dunning.ai, 93WoDuxx}, using molecular orbitals obtained from state-averaged complete active space self-consistent field (CASSCF) calculations. Under $\rm C_{2V}$ point group symmetry all \ai\ calculations started with  14 (8{}$\sigma$, 3{}$\pi_{x}$,  3{}$\pi_{y}$) orbitals which included 6 closed  (4{}$\sigma$, 1{}$\pi_{x}$,  1{}$\pi_{y}$) orbitals. The active space occupying 12 active electrons consisted of 8 ($5\sigma$--$8\sigma$,$2\pi, 3\pi$) valence orbitals. The PECs, including the 8 bound states \Xstate, \astate, \bstate, \cstate, \Aprimeprimestate, \Aprimestate, \Astate, and \Bstate\ are shown in Fig.~\ref{fig:ai_PECs}, as well as the adiabatic \estate\ - $(3)^1\Pi$ and \Cstate\ - \Cprimestate\ systems. The EAMC, SOCs (both diagonal and non-diagonal),  DMCs (diagonal and  transition) computed and used later for cross-section calculations are shown in Figs.~\ref{fig:ai_LSZ} -- \ref{fig:ai_EAMCs} in the original adiabatic representation as computed by \molpro\ 2015 \citep{MOLPRO2015} (left) and the diabatic representation (right). Further discussion of the diabatisation is given in the next section. 
%\green{I note that both diagonal and off-diagonal spin-orbit and dipole moment coupling curves are computed and are presented in figures \ref{fig:ai_LSZ} and \ref{fig:ai_DMZ}.}

\section{Diabatisation}
\label{sec:Diabatisation}

\subsection{Non-adiabatic couplings (NACs)}
\label{subsec:NACs}

The commonly employed Born-Oppenheimer approximation assumes the nuclear dynamics evolve adiabatically on a single, uncoupled electronic potential energy surface. This results in the appearance of avoided crossings between electronic states that are degenerate in energy at some geometry and large gradients in couplings connecting these states. The Born-Huang expansion \citep{88BoHuxx.book} allows one to recover the non-adiabatic behaviour in the region of the avoided crossing by including non-adiabatic couplings (NACs) which introduce off-diagonal kinetic energy terms in the molecular Hamiltonian. An alternative approach is to transform the electronic Hamiltonian to the diabatic basis, where electronic states are coupled via off-diagonal potential energy terms, termed diabatic couplings (DCs).\citep{81Delos.diabat, 82MeTrxx.diabat, 04JaKeMe.diabat}
The transformation from the adiabatic to the diabatic basis is described by a unitary matrix $U$, which is parametrically dependent on the NAC term,
\begin{gather}
{\bf U}(\beta(R))= \begin{bmatrix} \cos(\beta(R)) & -\sin(\beta(R)) \\ \sin(\beta(R)) & \cos(\beta(R)) \end{bmatrix},
\label{eq:U(beta)}
\end{gather}
where the mixing angle $\beta(R)$ is obtained by integrating the functional form of the non-adiabatic derivative coupling $\phi_{12}(R) := \bra{\psi_1}\frac{d}{dR}\ket{\psi_2}$  \citep{99SiHaWe.diabat, 15AnBaxx.diabat, 17BaAnxx.diabat,18KaBeVa.diabat}
\begin{equation}
\label{eq:beta(R)}
\beta(R)= \int^{R}_{-\infty} \phi_{12}(R') dR' 
\end{equation}
with $\ket{\psi_1}$ and $\ket{\psi_2}$ as the lower and upper energy electronic wavefunctions in the adiabatic representation.

Writing the two-state electronic Hamiltonian in terms of the adiabatic potential energy curves $V^{\rm a}_1(R)$ and $V^{\rm a}_2(R)$,
\begin{equation}
{\bf V}^{\rm a}(R)= 
\begin{pmatrix}
\centering
V^{\rm a}_1(R) & 0\\
0 & V^{\rm a}_2(R)
\end{pmatrix},
\end{equation}
One can obtain the diabatic Hamiltonian by applying the unitary transformation ${\bf U}(\beta(R))$,
\begin{multline}
{\bf V}^{\rm d}(R) = {\bf U^\dagger}{\bf V}^{\rm a}(R){\bf U} 
= \begin{pmatrix}
\centering
V^{\rm d}_1(R) & V^{\rm d}_{12}(R)\\
V^{\rm d}_{12}(R) & V^{\rm d}_2(R)
\end{pmatrix}
= \\
\begin{bmatrix} V^{\rm a}_1 \cos^2\beta + V^{\rm a}_2 \sin^2\beta    & \frac{1}{2}  (V^{\rm a}_{2}-V^{\rm a}_{1})  \sin(2\beta) \\  \frac{1}{2} (V^{\rm a}_{2}-V^{\rm a}_{1}) \sin(2\beta) &  V^{\rm a}_1 \sin^2\beta + V^{\rm a}_2 \cos^2\beta
\label{eq:dia_V}
\end{bmatrix},
\end{multline}
where the superscripts `d' and `a' refer to the diabatic and adiabatic bases, respectively, and the off-diagonal elements $V^{\rm d}_{12}(R)$ are the DCs. The reverse transformation is obtained by diagonalising the diabatic representation of the electronic Hamiltonian.

The NAC can be computed via quantum-chemistry methods from the electronic wavefunctions, as done by \citet{19SaNaxx.SO} for SO. It is symmetrical with a cusp  at the crossing point $R_{\rm c}$. Alternatively, the NAC curves are modelled using, e.g. a Lorentzian,\citep{81WeMexx.diabat} as given by
\begin{equation}
\phi_{ij}(R) = \phi^{\rm Lo}_{ij}(R;\alpha,R_{\rm c}) = \frac{1}{2}\frac{\alpha}{1+\alpha^2(R-R_{\rm c})^2},
\label{eq:lorentzian}
\end{equation}
where $\alpha$ is the inverse half-width-at-half-maximum (HWHM), or a Laplacian 
\begin{equation}
\phi_{ij}(R)=\phi^{\rm La}_{ij}(R;\gamma, R_{\rm c}) = \frac{\pi}{4\gamma}\exp\left[-\frac{|R-R_{\rm c}|}{\gamma}\right],
\label{eq:laplacian}
\end{equation}
where $\gamma$ is a damping constant related to the HWHM, superscripts `Lo' and `La' mean Lorentzian and Laplacian respectively, and the normalisation \(\int_{-\infty}^{\infty} \phi_{12}(R) \,dR\) = $\pi/2$ is applied. Figure~\ref{fig:NAC_MixAng_comparison} illustrates the \Cstate\ --\Cprimestate\ NAC modelled in this work using a Lorentzian and Laplacian function. The mixing angle $\beta(R)$, determined through Eq.~\eqref{eq:beta(R)}, ranges from 0  to $\pi/2$  going through $\pi/4$ at the crossing point $R=R_{\rm c}$ can also be seen in  Fig.~\ref{fig:NAC_MixAng_comparison}.

The Lorentzian was shown to provide a good description of the \ai\ NACs around the crossing point  \citep{09Varandas.diabat, 11MoVaxx.diabat, 15AnBaxx.diabat, 17BaAnxx.diabat} (see Fig.~\ref{fig:NAC_MixAng_comparison}) but diverges  at large distances $R$  from $R_{\rm c}$ causing improper shaped diabatic PECs by decaying too slowly. \citep{15AnBaxx.diabat,17BaAnxx.diabat, 09Varandas.diabat, 11Varandas.diabat}
It has been discussed that some damping functions can be introduced to correct the Lorentzian's slow decay using properties such as  dipole moments, but determination of their fitting parameters is both difficult and requires extra calculations. \citep{09Varandas.diabat,81WeMexx.diabat} 
Laplacians underestimate NACs far in the wings and overestimate them near the crossing point. \citep{15AnBaxx.diabat} One can show that NACs have an overlap dependence on the internuclear separation, \citep{09Varandas.diabat,09Vaxxxx.diabat} $R$, which is given properly by a Laplacian. \citep{15AnBaxx.diabat} The undesirable features of these NAC models can be mitigated by their combination \citep{15AnBaxx.diabat,17BaAnxx.diabat, 09Varandas.diabat, 11Varandas.diabat} to which we base our diabatisation procedure on. Our method of augmenting the Lorentzian with a Laplacian is discussed in section \ref{subsec:Diabatisation}.

\subsection{Diabatisation}
\label{subsec:Diabatisation}

Here we explore the so-called `property based diabatisation' method \citep{18KaBeVa.diabat} and construct diabatic potentials using the condition of having no avoiding crossing, which we define as to minimise their second derivatives in the vicinity of the crossing point $R_{\rm c}$:
\begin{equation}
\mathcal{F}=\sum_R \frac{d^2V_i^{\rm d}(R)}{dR^2} \to 0,
\label{eq:property_based_condition}
\end{equation}
hence creating the smoothest PECs $V^{\rm d}_1(R)$ and $V^{\rm d}_2(R)$. 

\begin{figure}[h!]
    \centering
    \includegraphics[width=0.48\linewidth]{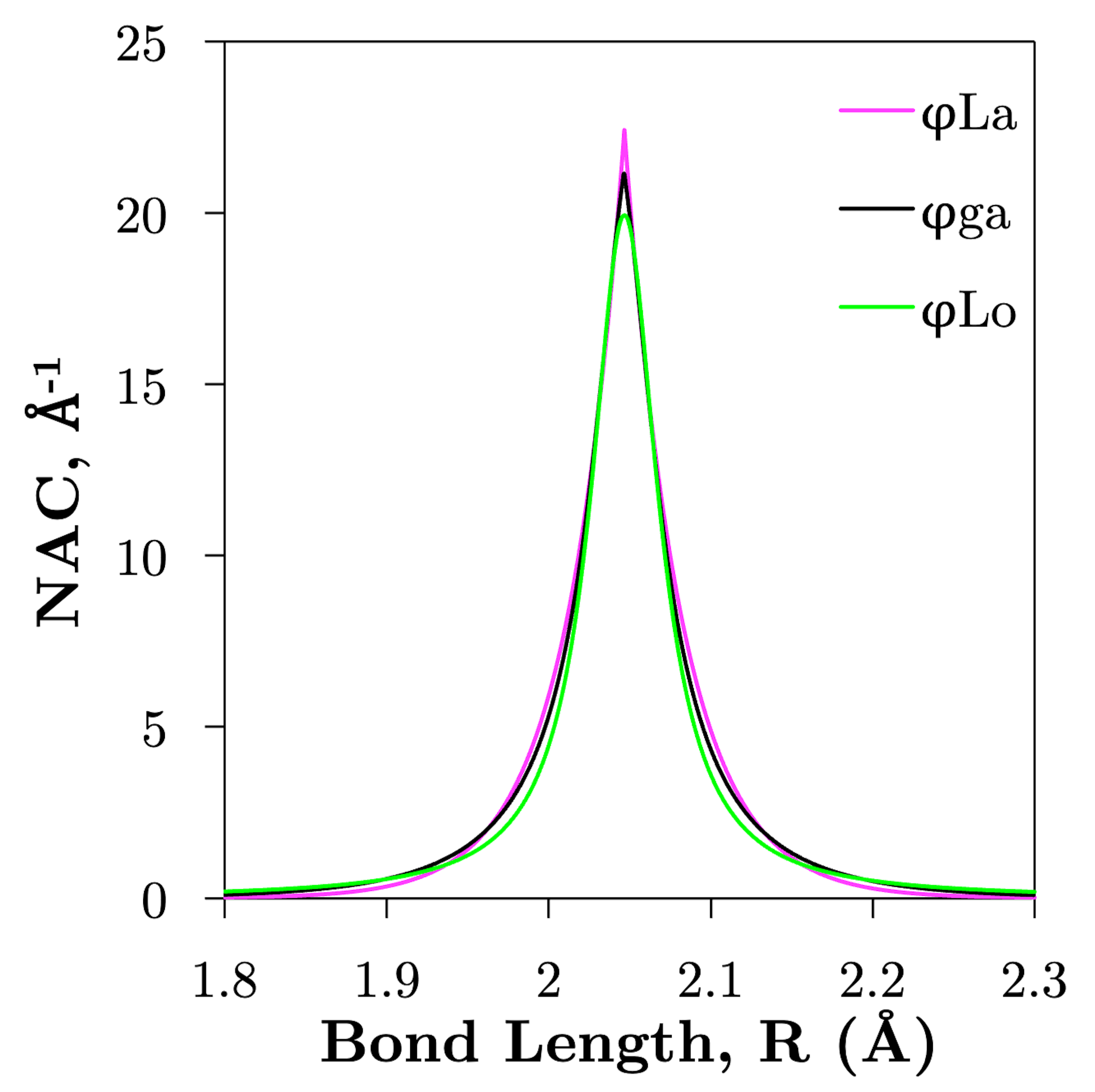}
    \includegraphics[width=0.48\linewidth]{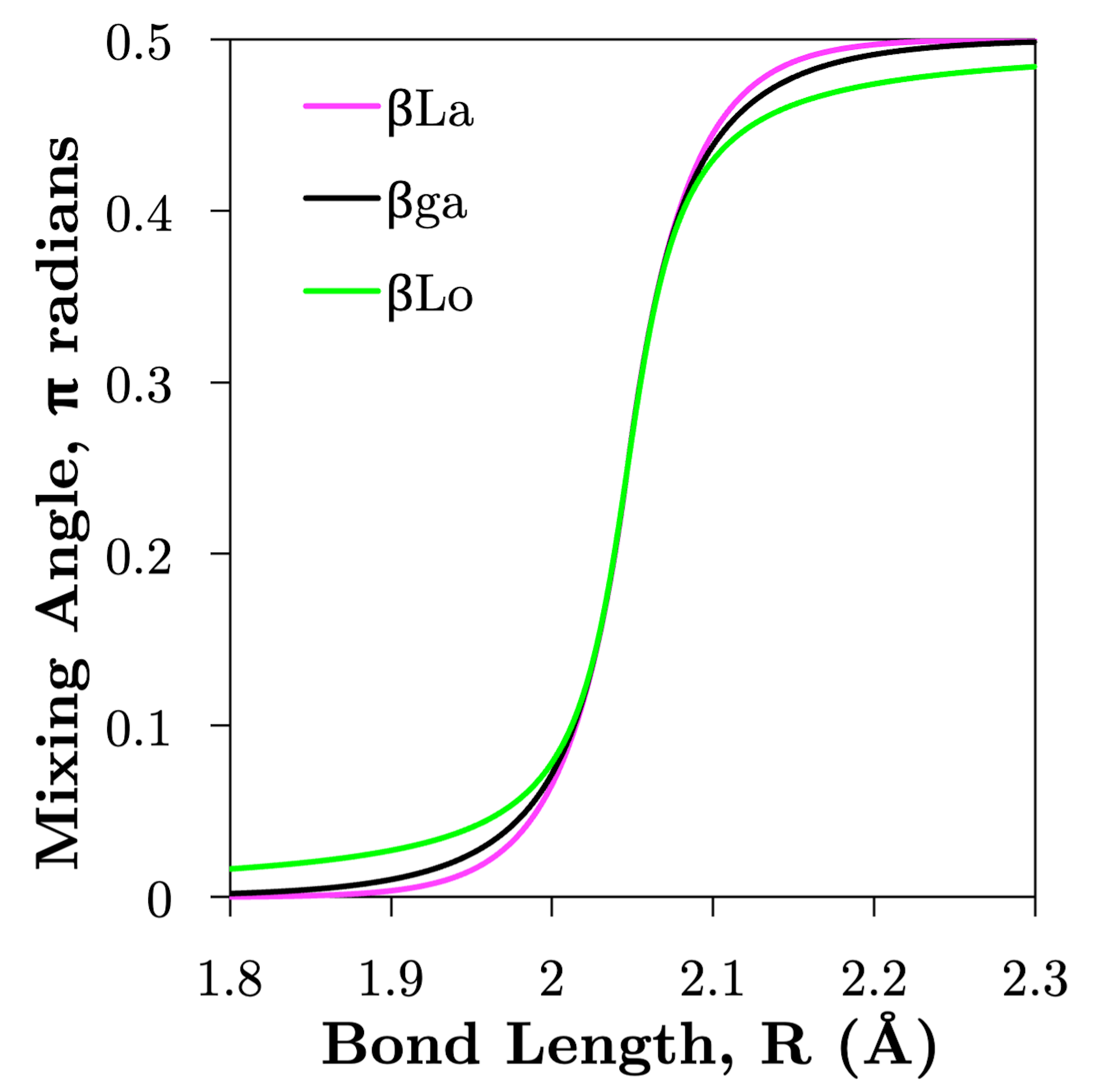}
    \caption{Comparison of example NACs ($\phi_{ij}$) and corresponding diabatic mixing angles ($\beta$) between the Lorentzian (`Lo'), Laplacian (`La'), and geometrically averaged (`ga') cases as described in the text. These curves are computed for the \Cstate\ and \Cprimestate\ non-adiabatic interaction (see section \ref{subsubsec:PEC} and Fig.~\ref{fig:ai_PECs}).}
    \label{fig:NAC_MixAng_comparison}
\end{figure}

In order to mitigate the undesirable properties of the Lorentzian and Laplacian functional forms, we follow the approach by \citet{15AnBaxx.diabat} and represent the mixing angle $\beta$ by the following combination of the mixing angles determined from the Lorentzian and Laplacian NACs in Eq.(\ref{eq:beta(R)}), $\beta^{\rm Lo}(R)$ and $\beta^{\rm La}(R)$
\begin{equation}
    \beta^{\rm ga}_{ij}(R)=\frac{1}{2}\arcsin{\left[\sqrt{\sin(2\beta^{\rm Lo}_{ij}(R))\sin(2\beta^{\rm La}_{ij}(R))}\right]},
\label{eq:beta_ga}
\end{equation}
where the `ga' superscript refers to the geometrically averaged diabatic mixing angle (See Fig.~\ref{fig:NAC_MixAng_comparison}). Equation \eqref{eq:beta_ga} must not be taken as the geometric average of $\beta^{\rm Lo}$ and $\beta^{\rm La}$, but rather originates from the simple geometric average of $V^{\rm Lo}_{12}$ and $V^{\rm La}_{12}$ (see \citet{15AnBaxx.diabat} and Appendix). \citet{15AnBaxx.diabat} also showed that an optimal relation between the parameters $\alpha$ and $\gamma$ exists which given by 
\begin{equation}
    \alpha \times \gamma = 1.397
\label{eq:alpha gamma relation}
\end{equation}
providing maximal overlap between the Lorentzian and Laplacian functions over the bond length.

\begin{figure}[h!]
    \centering
	\includegraphics[width=0.49\linewidth]{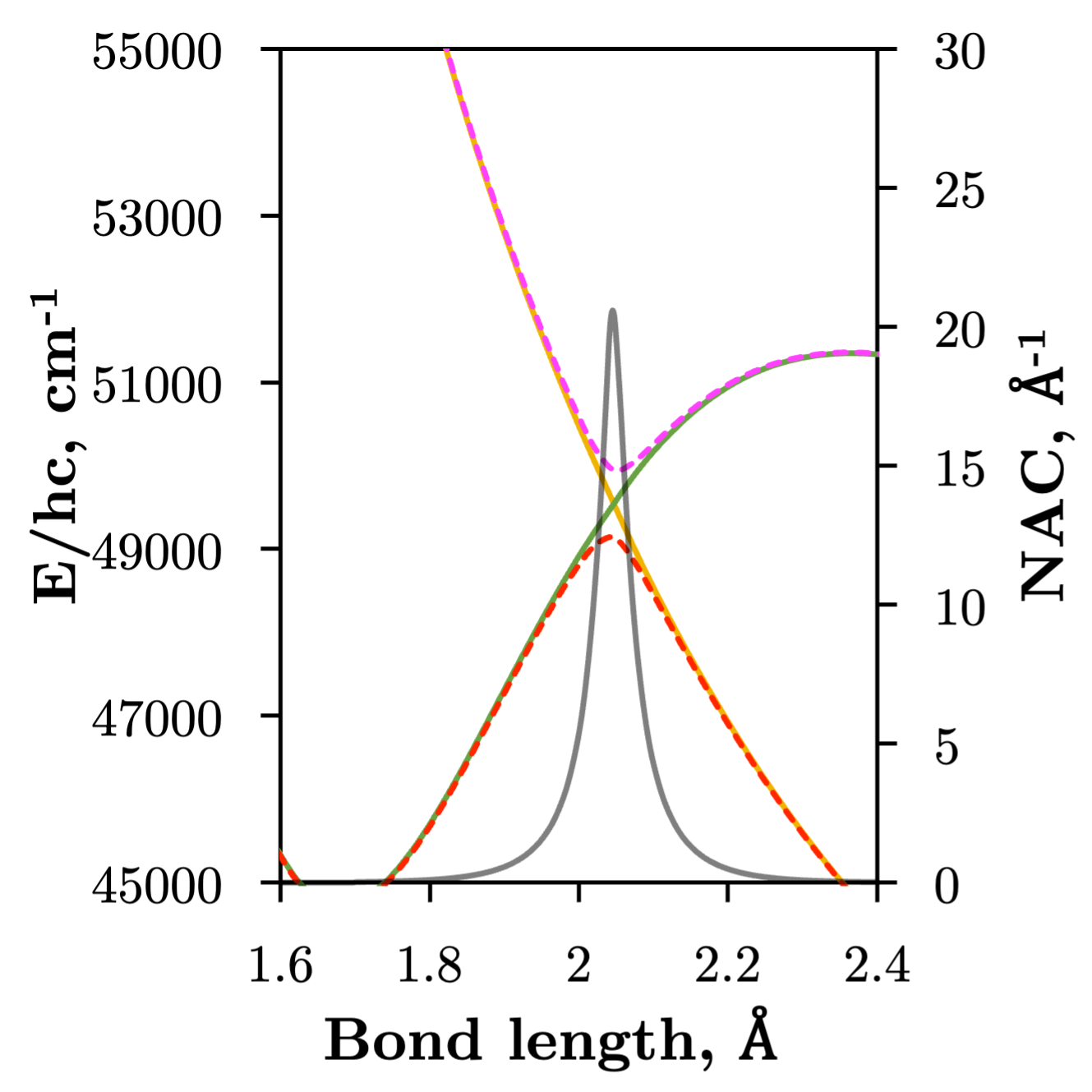}
	\includegraphics[width=0.49\linewidth]{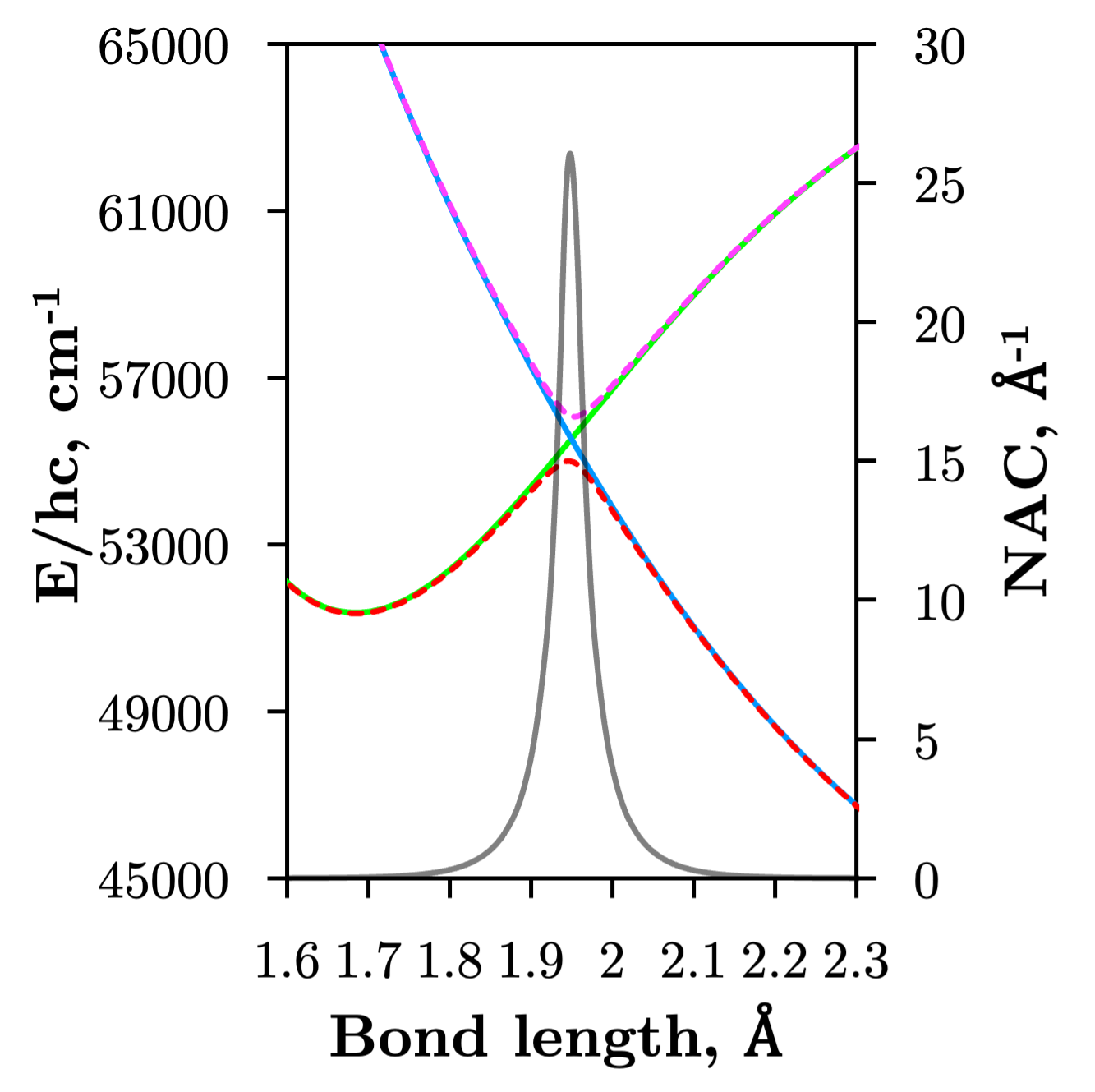}
	\includegraphics[width=0.49\linewidth]{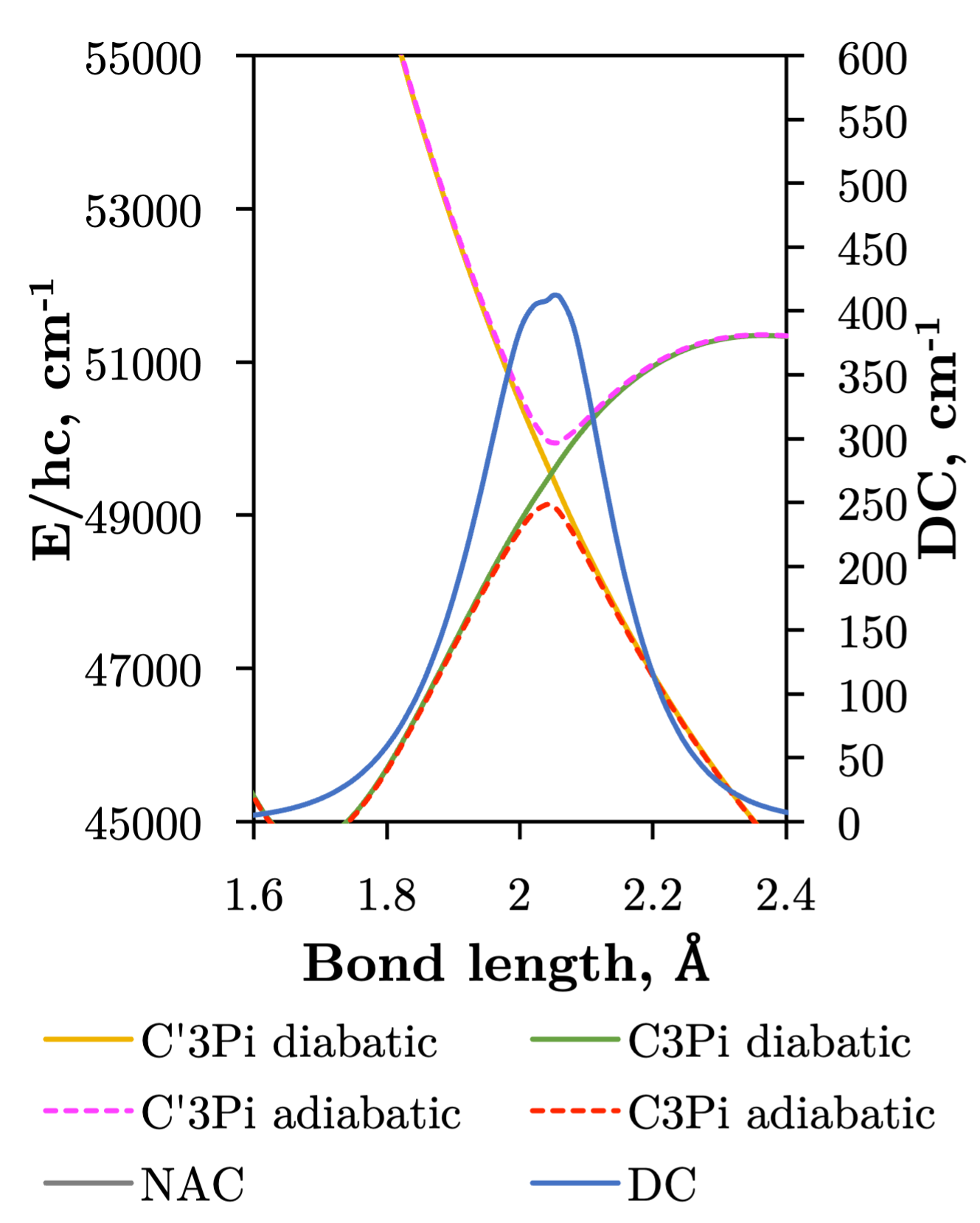}
	\includegraphics[width=0.49\linewidth]{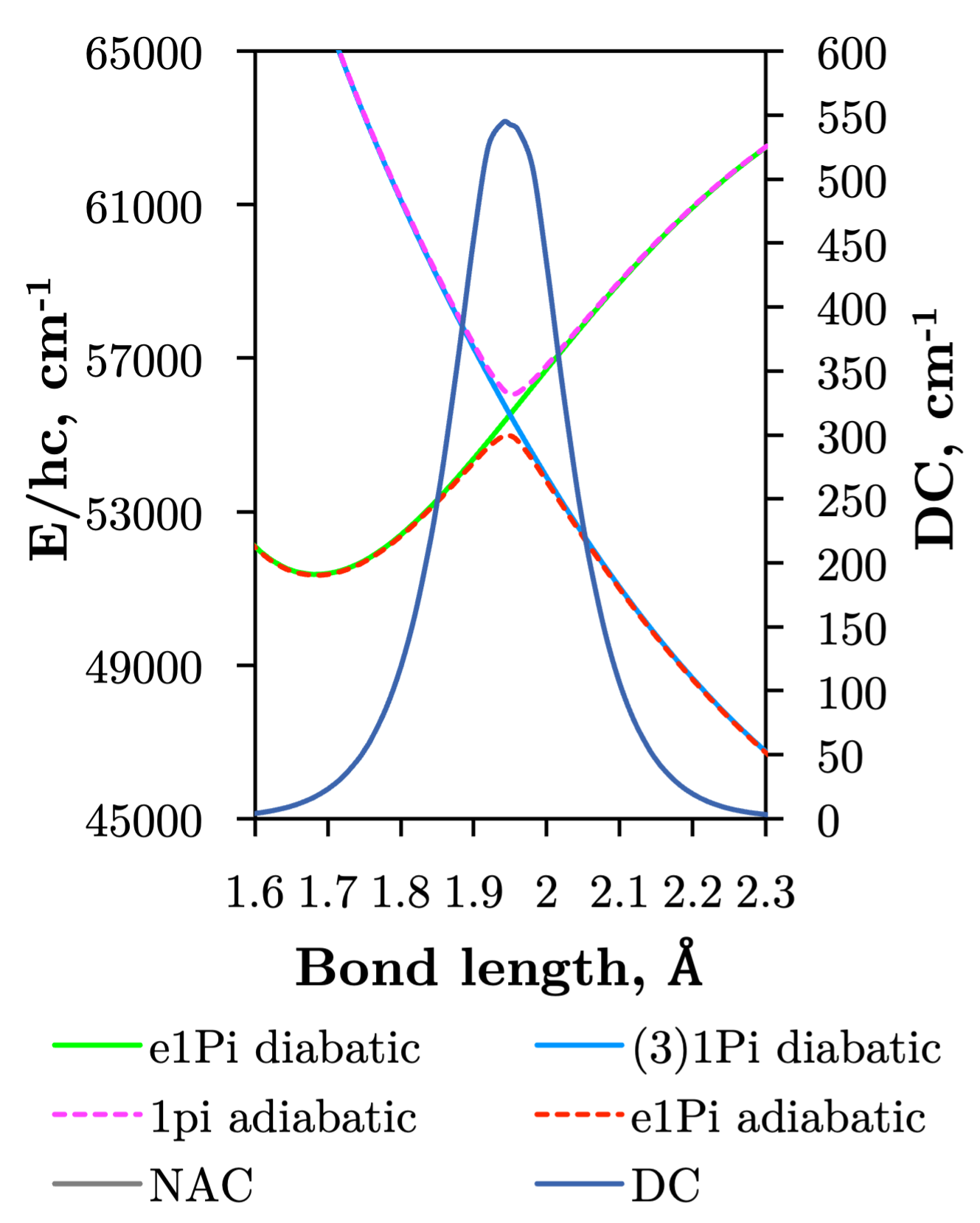}
    \caption{Illustration of the avoided crossings between  \estate\ - $(3)^1\Pi$ and (right panels) and \Cstate\ - \Cprimestate\ states of SO (left panels) are shown, where adiabatic PECs are presented as dashed lines and diabatic ones in solid lines. Superimposed are the DCs ($V^{\rm ga}_{12}$, bottom panels), and NACs ($\phi^{\rm ga}_{ij}$, top panels).}
    \label{fig:DCs}
\end{figure}

Where our method diverges from that of \citet{15AnBaxx.diabat} is in the determination of the crossing point $R_{\rm c}$ and the Lorentzian parameter $\alpha$. \citet{15AnBaxx.diabat} obtained $R_{\rm c}$ and $\alpha$ through fitting a Lorentzian to a NAC computed with \molpro. Instead, we determine a set of parameters $\{R_{\rm c}, \alpha \}$ by fulfilling the condition given in Eq.~\eqref{eq:property_based_condition}, to which the Laplacian parameter $\gamma$ is instantly obtained through Eq.~\eqref{eq:alpha gamma relation}. Using the theory developed in section \ref{subsec:NACs} and Eq.~(\ref{eq:beta_ga}) the diabatising transformation $\boldsymbol{U}^{\rm ga}$ corresponding to the `geometrically averaged' NAC is found. With this the diabatic potential energies and DC elements can be obtained through the simple matrix transformation in Eq.(\ref{eq:dia_V}). The diabatic PECs for SO can be seen in Fig.~\ref{fig:ai_PECs} and a closeup of the avoided crossings between the \estate\ - $(3)^1\Pi$ and \Cstate\ - \Cprimestate\ states of SO superimposed with their DCs, $V^{\rm ga}_{ij}$, and NACs, $\phi^{\rm ga}_{ij}$, are illustrated in Fig.~\ref{fig:DCs}. Figure \ref{fig:DCs} shows that the pair of singlet states are coupled more strongly than the triplet states, and also reveals the DCs to be slightly asymmetric. This is to be expected since the DCs depend on the difference $V^{\rm a}_2-V^{\rm a}_1$ which can be asymmetrical in nature. We see an effect of this especially in the DC between the triplet states where the adiabatic PEC turning points are offset to each other by $\sim 0.01$~\AA.

\subsubsection{Potential Energy Curves}
\label{subsubsec:PEC}

Figure~\ref{fig:ai_PECs} presents \ai\ PECs of the 13 lowest energy electronic states of SO. The \Cstate\ exhibits an avoided crossing at
$R$ $\sim$ 2.05~\AA\ due to a non-adiabatic coupling with the dissociative \Cprimestate, which lends a dissociative character to the \Cstate\ in the long-range region. Similarly, the \estate\ state exhibits an avoided crossing at 1.95~\AA, due to the singlet state $(3)^1\Pi$.\citep{11YuBixx.SO} These non-adiabatic interactions produce large gradients in coupling curves connecting these states within the region of the avoided crossing, as shown for EAMCs, SOC, and DMCs in Figs.~\ref{fig:ai_LSZ}--\ref{fig:ai_EAMCs}. The equilibrium geometry of the \Cstate\ state also lies very close in energy to those of the \dstate\ and \Bstate\ states, and so we can expect perturbations in the energy levels around their minima which was reported and confirmed experimentally by \citet{06LiElWe.SO} It is worth noting that the  \Aprimestate\ and \Aprimeprimestate\ states 
lie very close  across the range of nuclear geometries included in 
these calculations, but do not cross. 
 
Table \ref{tab:NAC params} provides values for the optimized NAC parameters $\alpha, \gamma, R_{\rm c}$ used to diabatise the energy degenerate pairs, which are visualised in Fig.~\ref{fig:DCs}. We see that the effect of diabatisation is to smooth the PECs, as enforced by Eq.\eqref{eq:property_based_condition} with no avoided-crossing. The non-Born-Oppenheimer dynamics, initially manifested in the nuclear kintetic energy, has been rotated into the potential and coupling terms connecting the non-adiabatically interacting states. We now produce PECs that can be easily modelled by analytical forms in the diabatic representation, allowing us to tune them to experimental data in a future study where we aim to produce an empirically accurate line list.

\begin{table}
\footnotesize
    \centering
    \caption{Optimised parameters $\alpha$ (inverse Lorentzian HWHM), $\gamma$ (Laplacian damping parameter), and $R_{\rm c}$ (avoided crossing/centroid position) for the Lorentzian and Laplacian NACs used to diabatise the \estate\ and \Cstate\ state PECs (see Fig.~\ref{fig:ai_PECs}).}
    \label{tab:NAC params}
    \begin{tabular}{lccc}
        \hline
        State & $\alpha$ & $\gamma$ & $R_{\rm c}$ (\AA) \\ 
        \hline\hline 
        \estate & 52.422 & 0.027 & 1.949 \\
        \Cstate & 39.859 & 0.035 & 2.047\\
        \hline
    \end{tabular}
\end{table}

Table \ref{tab:PEC-params} compares the equilibrium potential energies, $T_{\rm e}$ ($\rm cm^{-1}$), equilibrium bond lengths, $R_{\rm e}$ (\AA), and dissociation energies, $D_{\rm e}$ (eV), of the 11 lowest singlet and triplet states of SO  determined directly from the \ai\ adiabatic PECs presented in Fig.~\ref{fig:ai_PECs} to both calculations \citep{19SaNaxx.SO,11YuBixx.SO,00BoOrxx.SO, 00ArElWe.SO, 99BoOrxx.SO} and experiment.\citep{21BeJoLi.SO,79HeHuxx.book, 89Norwood.SO, 00ArElWe.SO, 05WaTaZh.SO} Our bond lengths show good agreement to both theoretical and empirical values, with better agreement to experiment than previous calculations for the \bstate, \Astate, and \dstate\ states. Worse agreements are seen for our $T_{\rm e}$ values, the most accurate being $T_{\rm e}(\Astate)$, $T_{\rm e}(\Cstate)$, and $T_{\rm e}(\estate)$ within \wav{152}, \wav{210}, and \wav{338} of experiment respectively. Lastly, we see worse agreements between our \ai\ dissociation energies to both experiment and calculations, to which we underestimate. It is apparent that our dissociation asymptotes are the major source of inaccuracy in our \ai\ SO model and probably arise because we do not include Sulfur specific diffuse functions in our basis set during \ai\ calculations. A future goal will be to mitigate the undesirable PEC features by refining our \ai\ model to experimental transition data, to which we will address when producing the final SO line list. We note that the reported \citet{21BeJoLi.SO} $R_{\rm e}$ values in Table \ref{tab:PEC-params} were derived from the $B_{\nu = 0 }$ rotational constant, and show close agreements to within \wav{0.006} \AA\ and \wav{0.002} \AA\ to our bond lengths for the \astate\ and \bstate\ respectively. Our fundamental vibrational energy  of the \Xstate\ state is found approximately \wav{30} too high from the experiment, which is to be expected with MRCI calculations, and provides insight to the accuracy of the other computed states and couplings, which will also require empirical tuning.

\begin{figure}[!ht]    
    \centering
    \includegraphics[width=\linewidth]{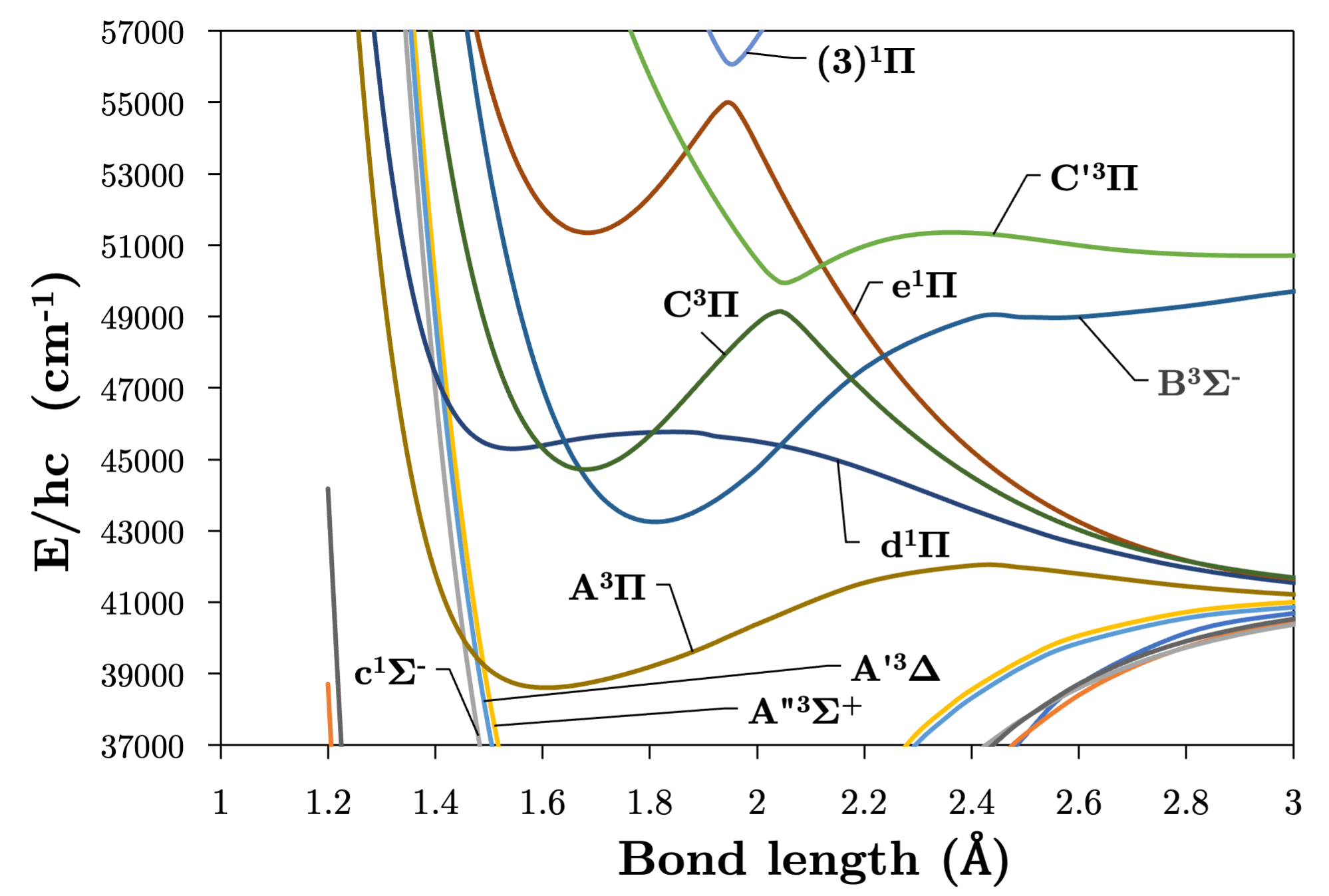}
    \caption{Our \ai\ PECs in the region 37,000 – 57,000 \cm\ showcasing the various state crossings.}
    \label{fig:ai_PEC_closeup}
\end{figure}

The intersections between states of different symmetries obtained in our calculations can be seen in Fig.~\ref{fig:ai_PEC_closeup}. The \dstate\ and \Cstate\ states cross at 1.59 and 1.82~\AA, \dstate\ and \Bstate\ at 1.64 and 2.20 \AA, both agreeing with \citet{11YuBixx.SO} intersection locations of 1.62 and 1.80 \AA\ and 1.60 and 2.14
\AA, respectively. The intersection of the \estate\ with the \Bstate\ state occurs at 2.4~\AA\ in our calculations, somewhat larger than the value of 2.3~\AA\ reported by \citet{11YuBixx.SO}. Since the \estate\ and \dstate\ states become repulsive at 1.92~\AA\ and 1.9~\AA\ respectively, crossings beyond these geometries provide potential predissociation pathways for the \Cstate\ and \Bstate\ states. \citet{11YuBixx.SO} also show that the \Cprimestate\ state crosses the \Bstate\ state at 2.25~\AA, to which they state predissociation of \Bstate\ through this \Cprimestate\ state is possible. \citet{19SaNaxx.SO} also give intersections $R(C,B)=1.57, 2.21$ \AA\ as opposed to our values of $R(C,B)=1.67, 2.18$ \AA. 
%\red{It is apparent from table \ref{tab:PEC-params} and our intersections that our computed \Bstate\ is reconstructed with the lowest accuracy out of the 11 states we consider. SHOULD WE SAY THIS?}

Lastly, we report further crossings between the \cstate, \Aprimestate, and \Aprimeprimestate\ states and the \Astate\ and \dstate\ states of SO at $R(c,A)$ = {1.46}~{\AA}, $R(A',A)$ = {1.48}~{\AA}, $R(A'',A)$ = {1.50}~{\AA}, $R(c,d)$ = {1.40}~{\AA}, $R(A',d)$ = {1.42}~{\AA}, and $R(A'',d)$ = {1.43}~{\AA}. These crossings agree with those reported by \citet{11YuBixx.SO} between \cstate, \Aprimestate, and \Aprimeprimestate with \Astate\ in the region {1.47}--{1.51}~{\AA} and crossings between \cstate, \Aprimestate, \Aprimeprimestate\ with \dstate\ in the region {1.42}--{1.45}~{\AA}.

\begin{table*}[h]
\footnotesize
    \centering
    \caption{Comparison of  \ai\ values of the equilibrium  potential energies $T_{\rm e}$ ($\rm cm^{-1}$), the dissociation energy $D_{\rm e}$ (eV) to the adiabatically correlated asymptotes, and equilibrium bond length $R_{\rm e}$ (\AA) from this work to the values from the literature. Parameters next to bold state symbols correspond to the \ai\ PECs calculated in this study.   Calculations \citep{11YuBixx.SO,00BoOrxx.SO, 00ArElWe.SO, 99BoOrxx.SO} use the MRCI/cc-aug-pV5Z  level of theory, calculation by \citet{19SaNaxx.SO} use the MRCI-F12+Q/cc-aug-pV(5+d)Z  level of theory, and experiments are  Photoion-Photoelectron Coincidence, \citep{89Norwood.SO} Multiphoton Ionization\citep{00ArElWe.SO} and $\rm Ar+SO_2$ afterglow, \citep{05WaTaZh.SO} spectroscopes as well as from the recent analysis by  \citet{21BeJoLi.SO}.}
    \label{tab:PEC-params}

    \begin{tabular}{lSSS|lSSS}
        \hline
        \multicolumn{1}{l}{State} & \multicolumn{1}{l}{$T_{\rm e}$ ($\rm cm^{-1}$)} & \multicolumn{1}{r}{$D_{\rm e}$ (eV)} & \multicolumn{1}{r}{$R_{\rm e}$ (\AA)} & \multicolumn{1}{|l}{State}  &  \multicolumn{1}{l}{$T_{\rm e}$ ($\rm cm^{-1}$)} & \multicolumn{1}{r}{$D_{\rm e}$ (eV)} & \multicolumn{1}{r}{$R_{\rm e}$ (\AA)} \\
        
        % State & $T_{\rm e}$ ($\rm cm^{-1}$) & $D_{\rm e}$ (eV) & $R_{\rm e}$ (\AA) & State & $T_{\rm e}$ ($\rm cm^{-1}$) & $D_{\rm e}$ (eV) & $R_{\rm e}$ (\AA) \\
        \hline\hline
$\boldsymbol{\Xstate}$	&	0	&	5.1253	&	1.4821	&	$\boldsymbol{\Aprimestate}$	&	29097.8878	&	1.503	&	1.7571	\\
Calc.\citep{11YuBixx.SO}	&	0	&	5.418	&	1.4865	&	Calc.\citep{11YuBixx.SO}	&	29828	&	1.72	&	1.7649	\\
Expt.\citep{79HeHuxx.book}	&		&	5.429	&	1.481	&	$\boldsymbol{\Bstate}$	&	43255.0097	&	0.9212	&	1.8121	\\
Calc.\citep{19SaNaxx.SO}	&	0	&	5.475	&   1.4925	&	Calc.\citep{19SaNaxx.SO}	& 41706.5886	&	1.4022	&	1.7868	\\
Expt.\citep{89Norwood.SO}	&		&		&	1.481	&	Calc.\citep{11YuBixx.SO}	&	41314	&	1.387	&	1.782	\\
Calc.\citep{00BoOrxx.SO}	&		&		&	1.481	&	Expt.\citep{89Norwood.SO}	&	41629	&	1.410\citep{13Rosen.book} &	1.775	\\
Calc.\citep{99BoOrxx.SO}	&		&		&	1.493	&	Calc.\citep{99BoOrxx.SO}	&	41206	&		&	1.794	\\
$\boldsymbol{\astate}$	&	5479.8013	&	4.4486	&	1.4979	&	$\boldsymbol{\dstate}$	&	45309.0766	&	0.0587	&	1.545	\\
Calc.\citep{11YuBixx.SO}	&	5936	&	4.682	&	1.4945	&	Calc.\citep{11YuBixx.SO}	&	44166	&	0.189	&	1.5475	\\
Expt.\citep{79HeHuxx.book}	&	5873	&	4.647	&	1.4919	&	Expt.\citep{00ArElWe.SO}	&	43902	&	0.195	&	1.5303	\\
Calc.\citep{99BoOrxx.SO}	&	5883	&		&	1.502	&	Calc.\citep{99BoOrxx.SO}	&	44975	&	0.059	&	1.723	\\
Expt. \citep{21BeJoLi.SO}	&		&		&	1.4920	&	Calc.\citep{00ArElWe.SO}	&	44471	&	0.14	&	1.553	\\
$\boldsymbol{\bstate}$	&	9774.1938	&	3.9154	&	1.5057	&	$\boldsymbol{\Aprimeprimestate}$	&	29731.2077	&	1.4417	&	1.765	\\
Calc.\citep{11YuBixx.SO}	&	10548	&	4.112	&	1.5062	&	Calc.\citep{11YuBixx.SO}	&	30495	&	1.637	&	1.7701	\\
Expt.\citep{79HeHuxx.book}	&	10510	&	4.137	&	1.5001	&	Expt.\citep{05WaTaZh.SO}	&	30692	&		&		\\
Calc.\citep{99BoOrxx.SO}	&	10576	&		&	1.514	&	Calc.\citep{99BoOrxx.SO}	&	30025	&		&	1.776	\\
Expt. \citep{21BeJoLi.SO}	&		&		&	1.5035	&	$\boldsymbol{\Cstate}$	&	44719.2593	&	0.5489	&	1.6786	\\
$\boldsymbol{\Astate}$	&	38607.6737	&	0.4246	&	1.6079	&	Calc.\citep{11YuBixx.SO}	&	44033	&	0.609	&	1.6692	\\
Calc.\citep{19SaNaxx.SO}	&	38879.2948	&  0.6441	&   1.5946	&	Calc.\citep{19SaNaxx.SO}	& 44909.0901	&	0.6027	&  1.6727 \\
Calc.\citep{11YuBixx.SO}	&	38334	&	0.665	&	1.6196	&	Expt.\citep{00ArElWe.SO}	&	44381	&		&	1.654	\\
Expt.\citep{79HeHuxx.book}	&	38455	&	0.662	&	1.609	&	Calc.\citep{99BoOrxx.SO}	&	44038	&		&	1.681	\\
Calc.\citep{00BoOrxx.SO}	&	38880	&		&	1.613	&	$\boldsymbol{\estate}$	&	51347.9346	&	0.4524	&	1.6864	\\
Calc.\citep{99BoOrxx.SO}	&	38931	&		&	1.719	&	Calc.\citep{11YuBixx.SO}	&	51224	&	0.45	&	1.6826	\\
$\boldsymbol{\cstate}$	&	27274.9752	&	1.7679	&	1.7571	&	Expt.\citep{00ArElWe.SO}	&	51558	&		&	1.6774	\\
Calc.\citep{11YuBixx.SO}	&	28544	&	1.879	&	1.7617	&	Calc.\citep{99BoOrxx.SO}	&	51258	&		&	1.685	\\
\hline
    \end{tabular}
\end{table*}

\subsubsection{Spin-Orbit and Electronic Angular Momentum Curves}
\label{subsubsec:SOCs}

Figure \ref{fig:ai_EAMCs} shows the EAMCs of SO both in the adiabatic and diabatic representations, where their relative phases are carefully maintained according with their \ai\ values \citep{14PaHiTe.AlO}. Without this, using couplings and any other non-diagonal properties in rovibronic calculations become meaningless. The 
phase of the EAMCs often changes after the crossing point, lending different long-range total angular momenta of the S+O atomic system important in dissociation mechanisms. 

\begin{figure}[h!]
    \centering
       \includegraphics[width=\linewidth]{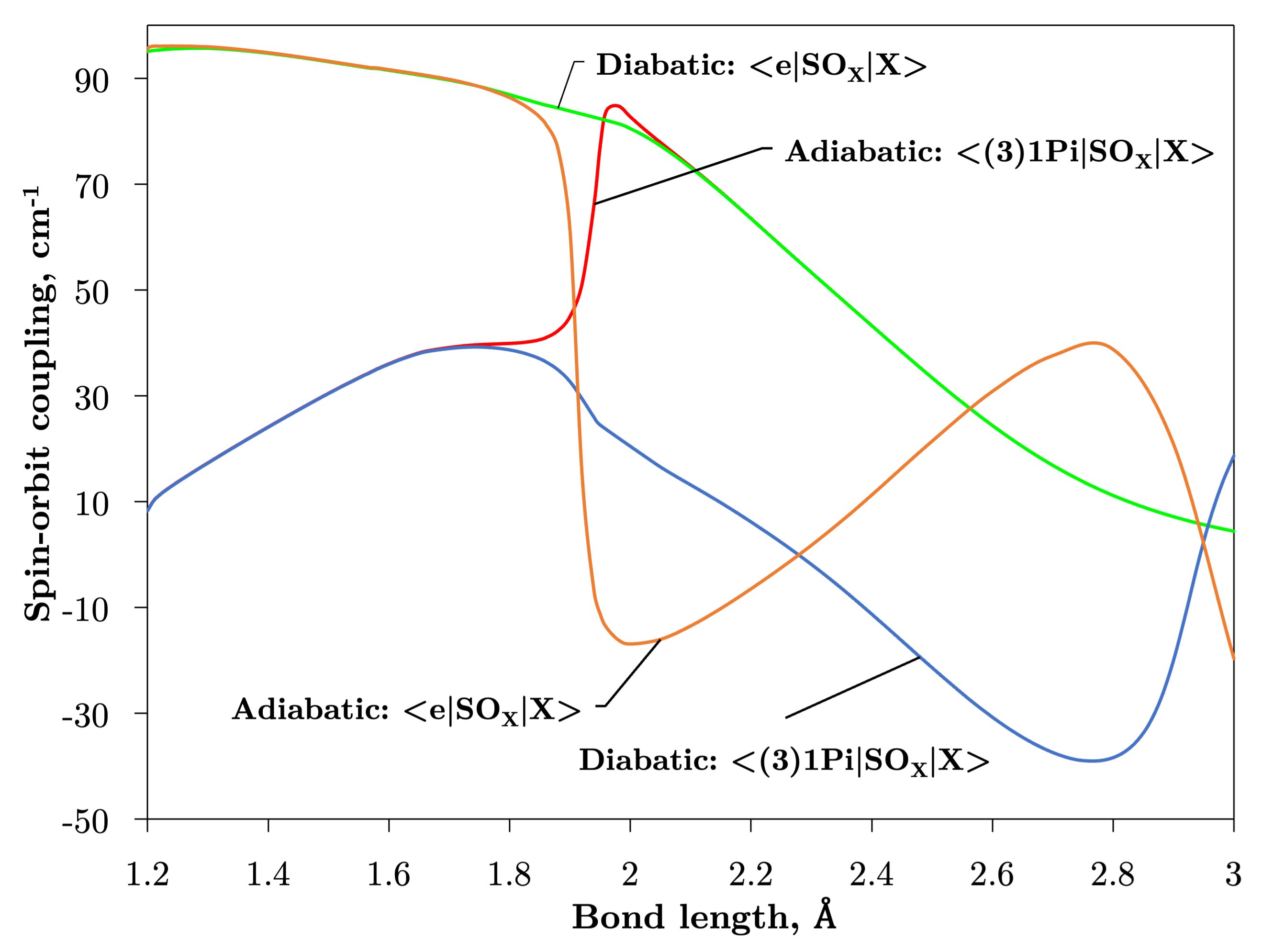}
    \caption{The diabatisation process for the curves $\bra{\estate}{\rm SO}_x\ket{\Xstate}$ and $\bra{(3)^1\Pi}{\rm SO}_x\ket{\Xstate}$, which become smooth as a result of the diabatisation. These curves have been multiplied by $-i$ for visual purposes.}
    \label{fig:ai_LSX:evolve}
\end{figure}

Figures \ref{fig:ai_LSZ} and \ref{fig:ai_LSX}  plot the $z$ (${\rm SO}_z$) and $x$ (${\rm SO}_x$) components of the spin-orbit curves of SO over nuclear geometries where the former couple states of same values of $\Lambda$ (projection of the electronic angular momentum)  and the latter couple states of different $\Lambda$. We see again that the effect of diabatisation is to smooth out the curves over $R$, where avoided crossings in the adiabatic picture create steep gradients around the avoided crossing. An example of the diabatisation process can be seen for the \brkt{\estate}{{\rm SO}_x}{\Xstate} and \brkt{(3)^1\Pi}{{\rm SO}_x}{\Xstate} pair in Fig.~\ref{fig:ai_LSX:evolve}. Spin-orbit matrix elements at the internuclear separation of the PEC crossings  are important in determining the possible spin-orbit induced predissociation mechanisms that occur between states of different spin multiplicity\citep{11YuBixx.SO} (see discussion below). Referring to the spin-orbit couplings \brkt{\estate}{{\rm SO}_x}{\Bstate}  and \brkt{\dstate}{{\rm SO}_x}{\Bstate} in  Fig. \ref{fig:ai_LSX} with the magnitudes of approximately 90 and {20}--{30}~{\cm}, respectively, the predissociation of the \Bstate\ state through \dstate\ is likely to be very weak, but will be stronger through the \estate\ state. The construction of diabatic SOCs will hence influence the efficiency of pre-dissociation pathways between states of different symmetry.

%The smoothening of $L_x$ curves wasn't done using our on-the-fly diabatisation procedure, but rather through extrapolating the curves to the atomic limit at large $R$. 

\begin{table*}[h!]
    \centering
    \caption{$\Sigma$ values (total spin angular momentum projection onto the internuclear axis) for the bra and ket electronic states of the SOCs presented in Figs. [\ref{fig:ai_LSZ}, \ref{fig:ai_LSX}]. }
    \label{tab:SOC Sigma}
    \begin{tabular}{ccccccccc}
        \hline
        SO coupling & bra $\Sigma$ & ket $\Sigma$ & SO coupling & bra $\Sigma$ & ket $\Sigma$ & SO coupling & bra $\Sigma$ & ket $\Sigma$  \\ 
        \hline\hline
\brkt{\Astate}{{\rm	SO}_x}{\Xstate}	&	0	&	1	&	\brkt{\estate}{{\rm	SO}_x}{\Xstate}	&	0	&	1	&	\brkt{\bstate}{{\rm	SO}_{z}}{\Bstate}	&	0	&	0	\\
\brkt{\Astate}{{\rm	SO}_x}{\Aprimestate}	&	0	&	1	&	\brkt{\estate}{{\rm	SO}_x}{\Bstate}	&	0	&	1	&	\brkt{\Aprimestate}{{\rm	SO}_{z}}{\Aprimestate}	&	1	&	1	\\
\brkt{\Cstate}{{\rm	SO}_x}{\Aprimestate}	&	0	&	1	&	\brkt{\Cstate}{{\rm	SO}_x}{\Aprimeprimestate}	&	0	&	1	&	\brkt{\Aprimeprimestate}{{\rm	SO}_{z}}{\Xstate}	&	1	&	1	\\
\brkt{\Astate}{{\rm	SO}_x}{\Bstate}	&	0	&	1	&	\brkt{\estate}{{\rm	SO}_x}{\Aprimestate}	&	0	&	1	&	\brkt{\Cstate}{{\rm	SO}_{z}}{\Astate}	&	1	&	1	\\
\brkt{\Cstate}{{\rm	SO}_x}{\Xstate}	&	0	&	1	&	\brkt{\dstate}{{\rm	SO}_x}{\Xstate}	&	0	&	1	&	\brkt{\Astate}{{\rm	SO}_{z}}{\Astate}	&	1	&	1	\\
\brkt{\Cstate}{{\rm	SO}_x}{\Bstate}	&	0	&	1	&	\brkt{\dstate}{{\rm	SO}_x}{\Bstate}	&	0	&	1	&	\brkt{\Aprimeprimestate}{{\rm	SO}_{z}}{\Bstate}	&	1	&	1	\\
\brkt{\Astate}{{\rm	SO}_x}{\Aprimeprimestate}	&	0	&	1	&	\brkt{\bstate}{{\rm	SO}_{z}}{\Xstate}	&	0	&	0	&	\brkt{\Cstate}{{\rm	SO}_{z}}{\Cstate}	&	1	&	1	\\
\brkt{\dstate}{{\rm	SO}_x}{\Aprimestate}	&	0	&	1	&	\brkt{\astate}{{\rm	SO}_{z}}{\Aprimestate}	&	0	&	0	&			&		&		\\
        \hline
    \end{tabular}
\end{table*}

%uodate plots

\subsubsection{Dipole Moment Curves}

Figures \ref{fig:ai_DMZ} and \ref{fig:ai_DMY}  plot the $z$- and $y$-components of the dipole moments coupling states of same and different symmetry ($\Lambda$ plus multiplicity) respectively.The corresponding $\mu_{x}$ components can be always reconstructed  from $\mu_{y}$  using their symmetry properties. We see again that the effect of diabatisation is to smooth out the curves over $R$, where now the DMCs tend to zero in the long range limit with no steep gradients caused by avoided crossings.

The vibronic intensities are directly affected by the derivatives of the dipole moment with respect to the internucelar separation, $R$. Since  adiabatic curves are prone to strong, steep-gradient variations around avoided crossings, even small inaccuracies in \ai\ calculations (including the position of the crossing and the corresponding NAC) can lead to large errors in spectral properties of the molecule. For example, the adiabatic \brkt{C^3\Pi}{\rm DM}{\Xstate}  dipole moment has a steep gradient at around 2~\AA\ which can be expected to be due to the avoiding crossing between \Cstate\ and \Cprimestate\ states, therefore the \Cstate--\Xstate\ electronic band is expected to be sensitive to the quality of its adiabatic description. The diabatic representation can also be sensitive to the quality of the corresponding curves, but to a significantly lesser extend due to their smooth character. 

Comparison with the \brkt{\Astate}{\mu_x}{\Xstate}, \brkt{\Cstate}{\mu_x}{\Xstate}, and \brkt{\Bstate}{\mu_z}{\Xstate} transition dipoles provided by \citet{19SaNaxx.SO} shows excellent agreements up to dissociation, with values (\{ours,\citet{19SaNaxx.SO}\}) at the ground state equilibrium geometry $R_e(\Xstate)=1.48$~{\AA} of $\{0.16,0.18\}$ $D$, $\{0.333,0.337\}$ $D$, $\{1.623,1.633\}$ $D$, respectively.

\subsection{Nuclear Motion Calculations}

\duo\ \citep{Duo} is a general purpose variational (open access\footnote{\href{https://github.com/Exomol/Duo)}{github.com/Exomol}}) code  that solves the rovibronic Schr\"{o}dinger equation for diatomics while allowing an arbitrary number of couplings between various electronic states including spin-spin, spin-orbit, spin-rotation, and rotational Born-Oppenheimer breakdown curves. It is assumed one has solved the Schr\"{o}dinger equation for the electronic motion \textit{a priori} in order to obtain PECs, SOCs, EAMCs, (T)DMCs etc. for the  electronic states  in question. These curves  can be supplied to the program  as either a grid of \ai\ points, or in an analytical form. After solving the Schr\"{o}dinger equation for the nuclear motion \duo\ obtains eigenstates and energies for the good quantum numbers $J$ (total angular momentum),  and $\tau$ (parity); other quantum numbers are assigned on the basis of the largest coefficient in the basis set. The eigenfunctions are used to compute transition line strengths and Einstein~$A$ coefficients in order to obtain a complete spectroscopy for the system in question. A detailed methodology of \duo\ is given by \citet{Duo}. 

%The coupled rovibronic quantum-mechanical Schr\"odinger equations are solved for the  electronic system of SO using the \duo\ program \citep{Duo}. The rovibronic eigenfunctions and eigenvalues are then used to generate a line list for SO. The  spectra of SO are computed using the ExoCross program \citep{ExoCross}.

\subsection{The \ai\ SO Spectrum}

Using the diabatic spectroscopic model we produce an \ai\ rovibronic spectrum of SO for the system involving the lowest 11 singlet and triplet electronic states of SO covering the wavelength range up to 147~nm. The ‘active’ couplings within
the spectrosocpic model used for cross-section calculations
include 23 SOCs, 23 (transition) DMCs, and 14 EAMCs in-line
with the couplings shown in Fig. \ref{fig:ai_LSZ}-\ref{fig:ai_EAMCs}. The vibrational sinc-DVR basis set was defined for a grid of 701 internuclear geometries in the range {0.6}--{6.0}~{\AA}. We select  58, 58, 49, 11, 31, 41, 27, 27, 14, 20, and 36 vibrational wavefunctions for the \Xstate, \astate, \bstate, \Astate, \Bstate, \cstate, \Aprimeprimestate, \Aprimestate, \Cstate, \dstate, and \estate, respectively, to form the contracted vibronic basis.  In total 15~364~624 Einstein A coefficients between 119~600 bound rovibronic states were computed 
with a maximum total rotational quantum number $J_\text{max}$ = 180 and used to simulate rovibronic absorption  spectra at a given temperature using the program \exocross.\citep{ExoCross}

%\green{We note that the \dstate\ vibrational basis far exceeds the dissociation limit for the state to avoid convergance issues, where experiment \citep{00ArElWe.SO} and calculations \citep{11YuBixx.SO,00ArElWe.SO} determine the maximum number of supported bound vibrational states to be $v=3$. Vertical structures appear in the \band{\dstate}{\Xstate} band of figure \ref{fig:All_comp_spec} which arise from resonances to un-bound \dstate\ states, a consequence of using a larger vibrational basis set. These continuum states will be filtered out in a future work producing the final line list.}

\begin{figure*}[h!]
    \centering
    \includegraphics[width=\linewidth]{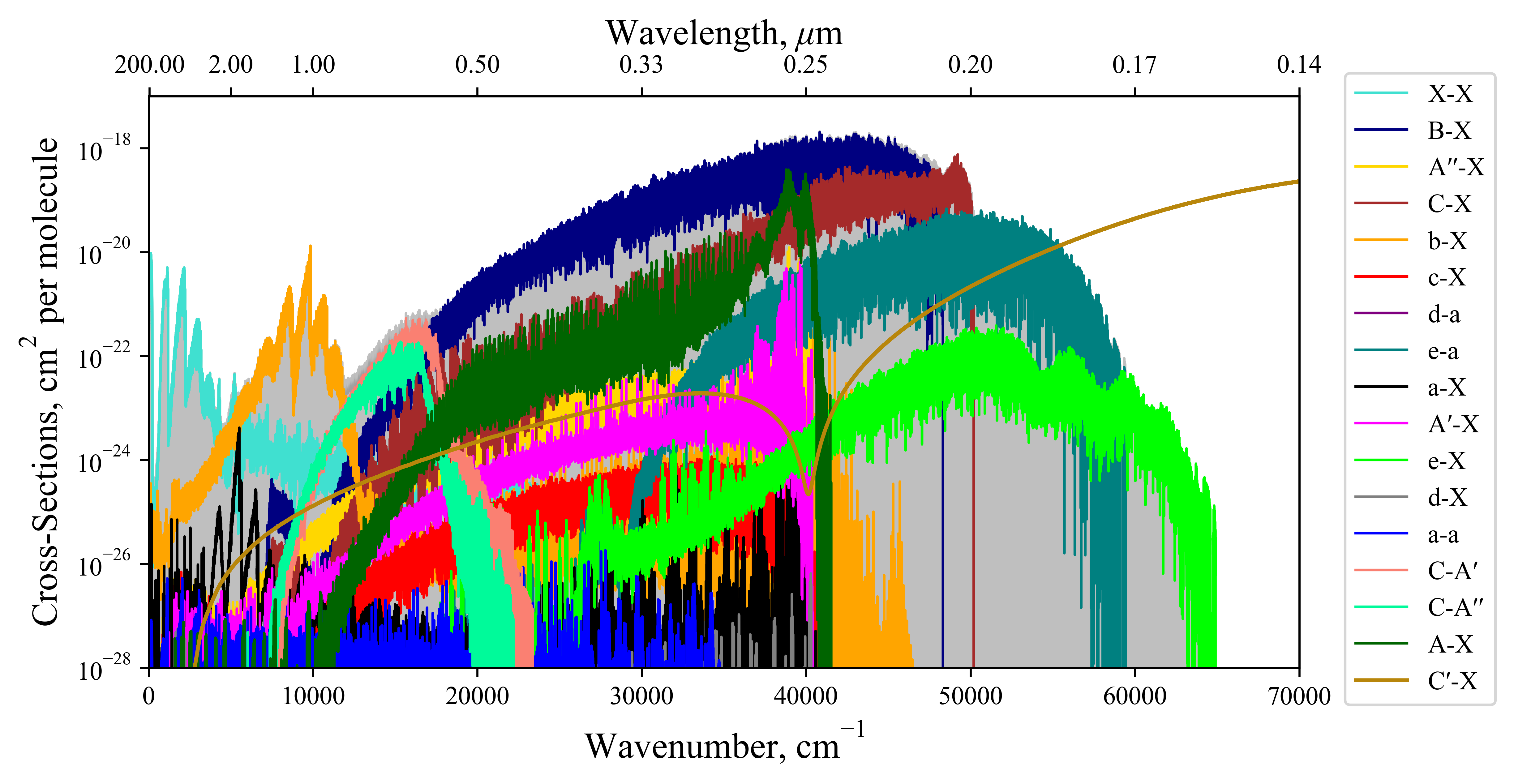}
    \caption{Dipole allowed and forbidden components of the \ai\ absorption spectrum simulated with the diabatic model at 5000~K connecting the \Xstate, \astate, \Aprimestate, and \Aprimeprimestate states. The \Cprimestate\ continuum is also plotted in gold. The absorption lines are modelled using Gaussian profiles with a HWHM of \wav{0.6} in the bound cases, and \wav{300} for the continuum band.}
    \label{fig:All_comp_spec}
\end{figure*}

Figure \ref{fig:All_comp_spec} shows an absorption rovibronic spectrum of SO computed at {5000}~{K} with all bands plotted using different colours, both electric dipole allowed and  forbidden.  We model the spectrum at a high temperature for a visual aid since at this temperature there is a good separation between different electronic bands. The grey shaded region in Fig.~\ref{fig:All_comp_spec} marks the total SO bound-bound absorption at {5000}~{K}, which is mostly traced by the strongest bands with the exception for the region between \wav{12000-17000}. Hence, weaker bands, e.g. ones that break dipole selection rules, have negligible contribution to the total SO opacity and will be less important for low resolution studies such as in astrophysical observations. 

The non-bound diabatic states such as \Cprimestate\ and $(3)^{1}\Pi$ are excluded from the bound-bound spectra simulations. Our tests show that the effect of the unbound states on the bound-bound spectra is negligible, and vice versa, the continuum spectra are negligibly affected by the bound electronic states and therefore can be treated separately.

The continuum spectra of the non-bound diabatic \Cprimestate\ state is computed using the method described in \citet{21PeYuTe} and is shown in figure \ref{fig:All_comp_spec}, plotted in gold and overlaying the bound-bound spectrum to demonstrate its contribution to the total SO opacity. For the continuum state a larger basis set of 5000 wavefunctions was used. The structure energetically above (below) the `dip' at \wav{41200} is due to absorption to unbound \Cprimestate\ states above (below) the $\rm S(^1D)+O(^3P)$ dissociation. The \band{\Xstate}{\Cprimestate} continuum band continues to \wav{100000}, peaking at $\sim$\wav{78000} which corresponds to the most vertical transitions from states localised around the minima of \Xstate.

We note that the dipole-forbidden bands in Fig.~\ref{fig:All_comp_spec} are not computed using quadrupole or magnetic dipole moments, which have very weak intensities, but rather intensities are `stolen' from other transitions. This intensity stealing propagates through the mixture of electronic wave-functions via couplings such as SOCs and EAMCs. For example, the spin-forbidden {\cstate}--{\Xstate} band occurs due to the overlap between the \cstate\ wavefunction  both with \estate\ and \dstate\ wavefunctions through the EAM couplings, and then with \Xstate\ through a secondary mixing via \brkt{\estate}{\rm SO_X}{\Xstate} and \brkt{\dstate}{\rm SO_X}{\Xstate} to produce a direct dipole moment, which dominates over the weaker magnetic and quadrupole moment mechanisms.

We compute absolute intensities for every rovibronic transitions between the lowest 11 diabatic singlet and triplet states of SO covering the entire spectroscopic range up to 147 nm, where NACs are treated. We note that the only other study with similar coverage into the UV on SO is from the theoretical work by \citet{19SaNaxx.SO} who compute cross sections for 190--300 nm. However, our spectroscopic model is both more complete and phase consistent (phases carefully reconstructed, see \ref{subsubsec:SOCs}), whereas \citet{19SaNaxx.SO} do not provide any phases.

\subsection{Experimental Coverage of the \ai\ SO Spectrum} 
\label{subsec:Exp_coverage}

\begin{figure}[h!]
    \centering
     \includegraphics[width=\linewidth]{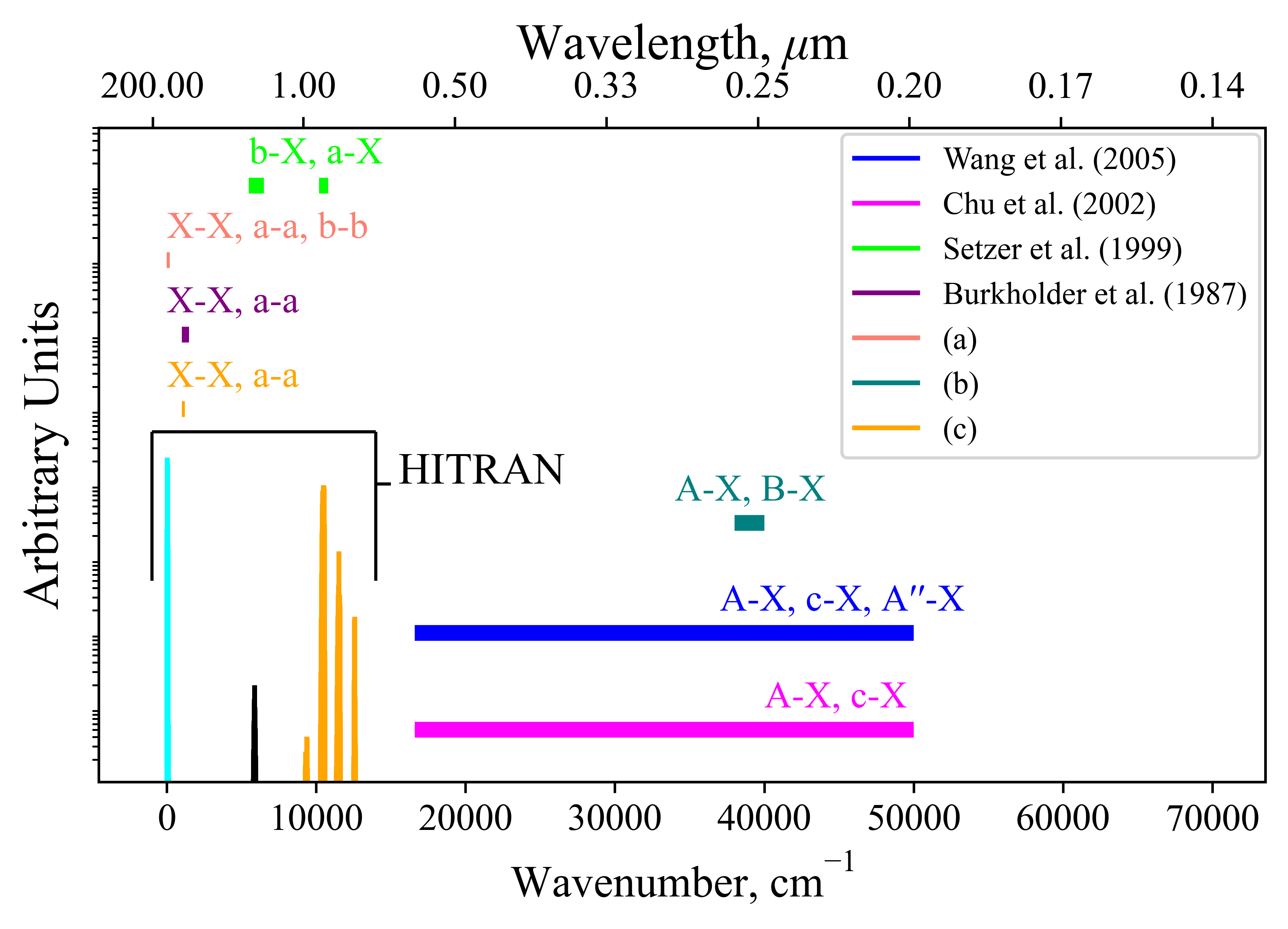}
    \caption{Coverage of experimental measurements for 24 sources  illustrated by horizontal bars covering the spectral regions measured, where the named works \citep{02ChWaLi.SO, 87BuLoHa.SO, 99SeFiRa.SO, 05WaTaHa.SO} include some spectral data, mostly with relative intensities: (a) 14 sources \citep{03KiYaxx.SO, 15MaHiMo.SO,17CaLaCo.SO,64PoLixx.SO, 64WiGoSa.SO, 74Tiemann.SO, 76ClDexx.SO, 82Tiemann.SO, 87EnKaHi.SO, 93Yamamoto.SO,94CaClCo.SO,96KlSaBe.SO,97BoCiDe.SO,97KlBeWi.SO,CDMS} cover \band{\Xstate}{\Xstate}, \band{\astate}{\astate}, \band{\bstate}{\bstate} for \wav{0-125}; (b) 3 experimental sources \citep{82Colin.SO,86ClTexx.SO, 94StCaPo.SO} cover the \band{\Astate}{\Xstate} and \band{\Bstate}{\Xstate} bands for \wav{38000--39800}; (c) 2 experimental sources \citep{85KaBuKa.SO,88KaTiHi.SO} cover the \band{\Xstate}{\Xstate} and \band{\astate}{\astate} bands for \wav{1040--1125}.}
    \label{fig:HITRAN-THeory_comp}
\end{figure}

Currently within the literature a small fraction of the SO spectrum has been measured experimentally covering only the \Xstate, \astate, \bstate, \Astate, and \Bstate\ states. Figure \ref{fig:HITRAN-THeory_comp} reviews the spectroscopic coverage of SO from 24 experimental sources from the literature. Figure \ref{fig:All_comp_spec} shows our model to supplement the experimental data over the whole spectral range. In particular, we cover the SO spectrum above \wav{40000} and \wav{12000--16000} where no measurements have been taken for any electronic state. We also plot the available  \hitran \citep{HITRAN2012} SO line list containing data on the first three electronic states \Xstate, \astate, and \bstate. Our \ai\ model is able to extend the \hitran\ coverage up to dissociation at \wav{40000}. 

The aim of a future work is to refine our \ai\ SO model to the experimental transition frequencies from these sources and to produce an empirically accurate line list for SO. 
%empirically derived energy levels obtained from experimental transition frequencies \citep{02ChWaLi.SO, 87BuLoHa.SO, 99SeFiRa.SO, 05WaTaHa.SO, 03KiYaxx.SO, 15MaHiMo.SO,17CaLaCo.SO,64PoLixx.SO, 64WiGoSa.SO, 74Tiemann.SO, 76ClDexx.SO, 82Tiemann.SO, 87EnKaHi.SO, 93Yamamoto.SO,94CaClCo.SO,96KlSaBe.SO,97BoCiDe.SO,97KlBeWi.SO,CDMS, 82Colin.SO,86ClTexx.SO, 94StCaPo.SO, 85KaBuKa.SO,88KaTiHi.SO} 

%In this future work we  compare our refined spectra to experimental measurements provided by \citet{87BuLoHa.SO, 99SeFiRa.SO, 05WaTaZh.SO, 02ChWaLi.SO} to assess the accuracy of the model. 
%The aforementioned sources only provide spectra in non-LTE conditions, meaning direct intensity comparisons are not possible. 
%From an initial analysis of the \ai\ SO spectrum, we were able to produce non-LTE models that reproduce the structure of experimental spectra well with the exception of shits between the line positions. These shifts are to be expected from the the accuracy of our MRCI calculations (see section \ref{subsubsec:PEC}). However, the aim of this work is not to attain ultimate accuracy within our model, an aim of this future SO study, but rather to provide a completely reproducible diabatic \ai\ model, complete spectroscopic coverage, and complete description of the electronic structure of SO. 

\section{Effect of diabatisation on the computed Spectra}
\label{sec:effect of dia}

In theory, the adiabatic and diabatic representations should provide identical  results, provided that the corresponding  NACs and DCs are included. However, due to the computational cost of computing NACs through proper \ai\ methods, it is not common practice to include NACs in adiabatic models. Without definition of the NAC, non-Born-Oppenheimer interactions are effectively removed. 

%\subsection{Diabatic States to the Continuum}
%\label{subsec:DCs}

In this section we analyse the importance of the non-adiabatic couplings between \Cstate\ and \Cprimestate\ as well as between  \estate\ and  $(3)^{1}\Pi$ for computing the (absorption) spectra of SO. To this end we consider our diabatic spectroscopic model to be complete and use it to compute a reference absorption spectrum of SO. This spectrum is then used to compare with an `adiabatic' spectrum of SO computed using the non-adiabatic curves \textit{without} NACs. 

\begin{figure}[h!]
    \centering
    \includegraphics[width=\linewidth]{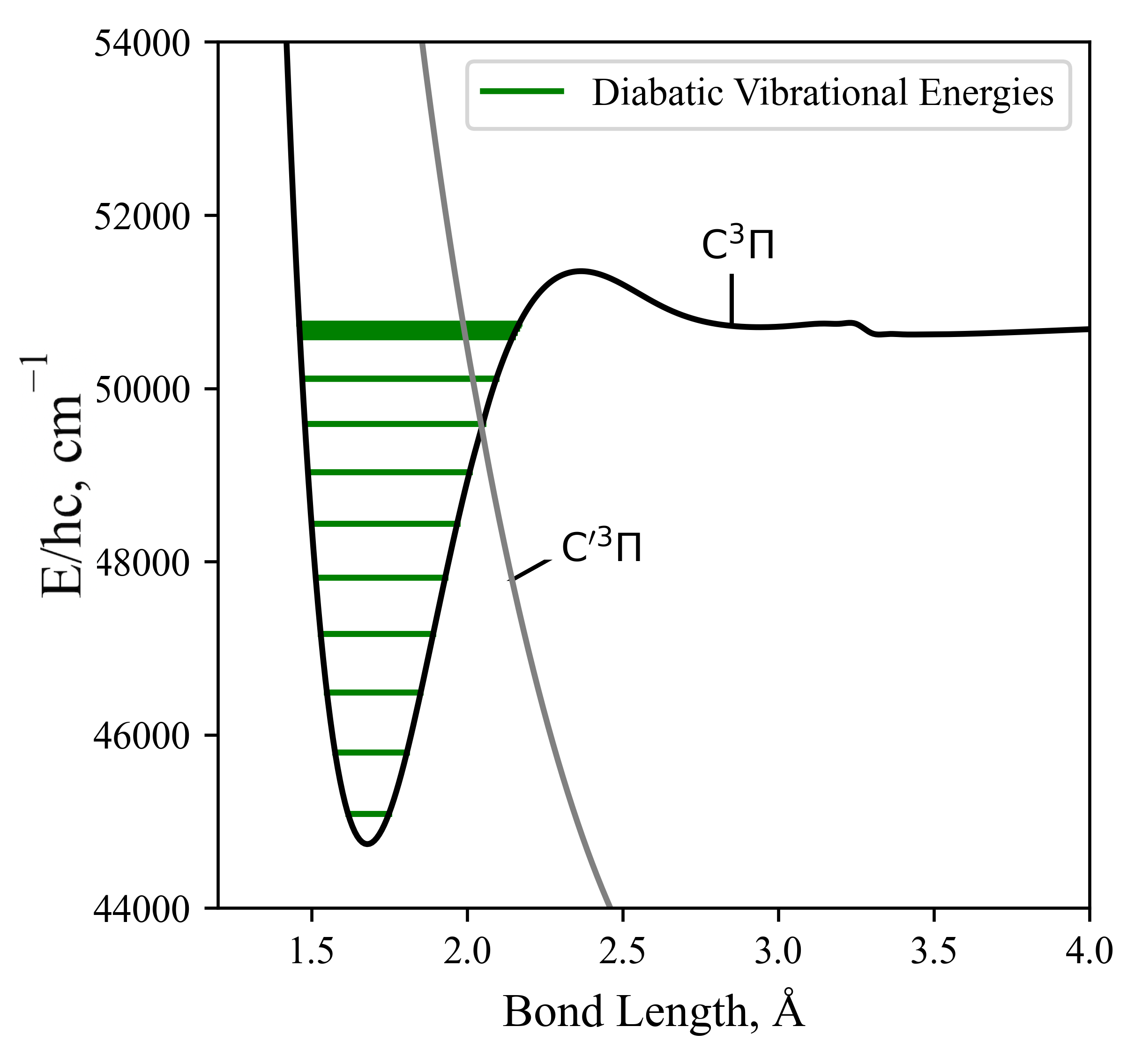}
    \caption{The computed \duo\ diabatic vibronic energies of the \Cstate\ superimposed upon the \Cstate\ and \Cprimestate\ PECs. The secondary and tertiary `bumps' at $\rm R = 2.42, 3.30$ \AA\ are due to an avoided crossing and numerical noise, respectively.}
    \label{fig:C-energies}
\end{figure}

As detailed in section \ref{subsec:NACs}, the diabatisation of the potential energy curves induces non-zero off-diagonal elements via corresponding  2$\times$2 potential energy matrices in the diabatic representation. These off-diagonal coupling elements will be referred to as diabatic couplings, DCs, which are characterised by cusp shaped curves centered at the avoided crossing. 
As before, in the diabatic spectrum simulations the unbound states are excluded. Figure \ref{fig:C-energies} illustrates the vibrational energy levels of the diabatic \Cstate, fully bound below its dissociation limit of \wav{50700}.

%Figure \ref{fig:C-energies} shows the case B \Cstate\ diabatic vibronic energy levels which are negligibly impacted by the inclusion of the DCs and \Cprimestate\ within the model. We calculate a mean energy level difference of $\overline{E^{\rm B}(\Cstate)-E^{\rm A}(\Cstate)}=-9.64\times 10^{-6}$ \cm\ and standard deviation of $\sigma(E^{\rm B}(\Cstate)-E^{\rm A}(\Cstate))=2.71\times 10^{-5} $ \cm. This result shows that the upper repulsive states and DCs have negligible impact on the physics associated with the lower bound diabatic counterparts and that the important dominating non-Born-Oppenheimer dynamics must be manifested within the morphology of the diabatic PECs. We aim to quantify this result in a future work, but here we conclude that the diabatic representation without definition of DCs and upper coupled states is likely more physical than an adiabatic model without treatment of NACs.

\begin{figure*}[ht]
    \centering
    \includegraphics[width=0.48\linewidth]{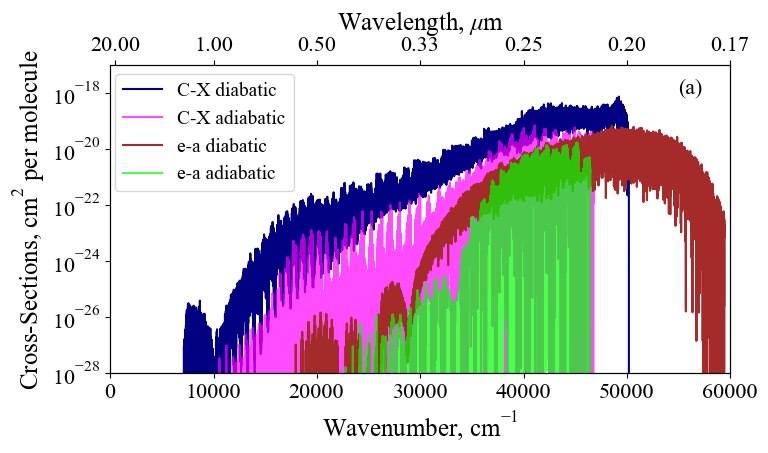}
    \includegraphics[width=0.48\linewidth]{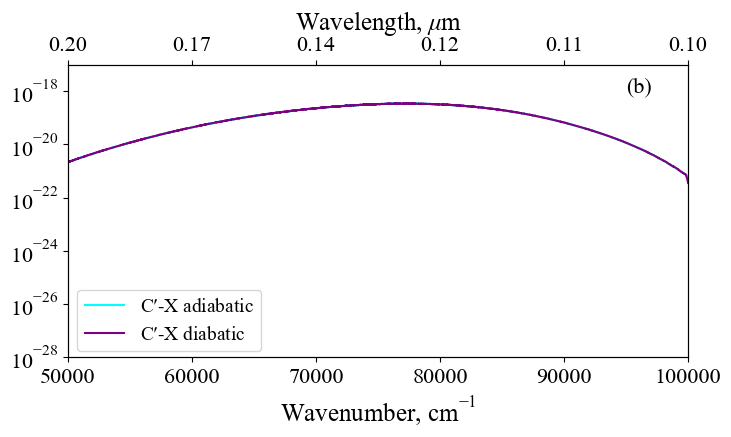}
    \caption{A comparison between the \band{\Xstate}{\Cstate}, \band{\astate}{\estate}, and \band{\Xstate}{\Cprimestate} band spectra computed with an adiabatic model with no defined NAC and a diabatic model. These bands are dipole allowed and are expected to be observable, we see great differences between the spectra at the dissociation, highlighting the importance of diabatisation. Each spectra has been modelled at a temperature of 5000~K with Guassian line profiles of a \wav{0.6} HWHM. %\red{Fixed micron scale and legend text size}
    }
    \label{fig:CX_ea_comp_Spec}
\end{figure*}

Figure \ref{fig:CX_ea_comp_Spec} illustrates the importance of the non-adiabatic effects when modeling the spectra around avoiding crossings for  the \band{\Xstate}{\Cstate} and \band{\astate}{\estate} bound-bound absorption bands (panel a), and the \band{\Xstate}{\Cprimestate} continuum absorption band (panel b).  
The adiabatic spectra were computed with the NACs excluded and compared to the diabatic spectra with  the non-adiabatic effects fully encountered.  Each spectra has been modelled at a temperature of 5000~K -- such that hot bands are populated, aiding our comparisons below -- with Gaussian profiles of a \wav{0.6} half-width-at-half-maximum (HWHM) for the bound-bound spectra and a HWHM of \wav{300} for continuum bands.

Great differences between the bound-bound spectra in panel (a) of Fig.~\ref{fig:CX_ea_comp_Spec} are seen towards both the high and low energy regions. In the high energy region the  adiabatic spectral bands terminate abruptly at the avoided crossings whereas the diabatic bands continue to the diabatically correlated dissociation asymptotes $\rm S(^1D)+O(^3P)$ \& $\rm S(^1D)+O(^1D)$ (see Fig.~\ref{fig:ai_PECs}). The diabatisation extends these bands by at least a few thousand wavenumbers because of the availability of higher rovibrational states in the deeper diabatic potential wells. For purely bound-bound calculations, the adiabatically computed bands have lower intensities compared to the diabatic spectrum which can be attributed to the increased repulsive character of the adiabatic PECs on the right hand side of the crossing points present. 
%character of the adiabatic \Cstate\ and \estate\  PECs means their wavefunctions have an unbound character for geometries beyond the avoided crossing. 
Due to the tunneling through the potential barriers, the adiabatic wavefunctions `leak' to the continuum region thus resulting  in reduction of the intensity of their bound absorption spectra. %In the case of the \band{\astate}{\estate} band, we see great overlap in cross sections between $\sim$\wav{35000-45000}  %We saw the FC factors to increase by a factor of $\sim 2$ and $\sim 4$ for the \band{\Xstate}{\Cstate} and \band{\astate}{\estate} systems, respectively. 
The most interesting feature from Fig.~\ref{fig:CX_ea_comp_Spec} is the extension of the \band{\astate}{\estate} band beyond the stronger \band{\Xstate}{\Cstate} band at $E/hc >$ \wav{50000}. Although being relatively weak, this band is not covered by stronger bands and therefore may be observable in the $\sim 0.18-0.2$~$\mu$m region, a result only predicted when using a full non-adiabatic theoretical treatment.

%\noindent to be small and within the error of an \ai\ model. However, this effect is greater within the \band{\Aprimestate}{\Cstate} and \band{\Aprimeprimestate}{\Cstate} bands which have less-vertical transitions than \band{\Xstate}{\Cstate} and \band{\astate}{\estate} where we see intensity differences of up to one order of magnitude. The most interesting feature from Fig.~\ref{fig:CX_ea_comp_Spec} is the extension of the \band{\astate}{\estate} beyond the stronger \band{\Xstate}{\Cstate} band at $E/hc \gtrsim$ \wav{50000}. Thus, in the adiabatic picture without treatment of NACs, one would expect this weaker band to be un-observable, but as revealed by the diabatisation \band{\astate}{\estate} extends into the high energy region where it isn't covered by stronger bands.

%\green{ The regions where the continuum and bound parts of the adiabatic spectra meet appear to be caused by transitions to states dominated by un-bound and bound characters for the \Cstate\ and \estate\ respectively. This causes the bands connecting the \Cstate\ to be more `continuous' at this boundary as opposed to the adiabatic \band{\astate}{\estate} band having a `dip' like structure where bound-bound transitions begin to dominate.}

%We note that these regions negligibly contribute to the total SO opacity and so are not important for the SO model, however we report them since these features will be important for other systems where non-adiabatic effects occur in the spectroscopically important regions. 
%These flat features are the corresponding continuum bands. 

The low intensity regions are very sensitive to changes in the \ai\ model and will be also affected by the changes in the shape of couplings between the adiabatic and diabatic representations. The hump-like structure in the {\Cstate}--{\Xstate} band at around \wav{10,000}  is absent in the adiabatic spectrum because of the unavailability of vibrational states above the avoided crossing. For example, the brightest transitions within this hump for the {\Cstate}--{\Xstate} band connect the $\nu =13$ state which is energetically above the avoided crossing in the adiabatic PEC.

% \green{If continuum absorption bands for the systems in Fig.~\ref{fig:CX_ea_comp_Spec}(a) were also computed the low energy region would extend almost to \wav{0} and be continuum in character as opposed to the diabatic counterparts terminating at higher wavenumbers. This difference is attributed to the unbound character of the adiabatic \estate\ and \Cstate\ PECs correlating to the lower $\rm S(^3P)+O(^3P)$ dissociation whereas the diabatic counterparts are bound with upper dissociation limits. This would be nonphysical since we know the diabatic picture provides more complete physics, highlighting the importance in the treatment of non-adiabatic effects.}

 We note that these regions negligibly contribute to the total SO opacity and so are not important for the SO model,  but will be important for other systems where non-adiabatic effects occur in the spectroscopically important regions.
 
%The continuum component of the adiabatic bands connect states of progressively high vibrational excitations in the lower state and low excitations within the upper electronic state, thus this continuum character becomes stronger at higher temperatures. From our calculations the continuum character of these adiabatic bands only exceed $10^{-28}$ $\rm cm^{-2}$ per molecule for temperatures above 3000 K.

Panel (b) of Fig.~\ref{fig:CX_ea_comp_Spec} presents a similar analysis for the continuum \band{\Xstate}{\Cprimestate} band, which would include an additional bound structure towards longer wavelengths if the NAC is not included in the adiabatic model since the \Cprimestate\ PEC in this representation is bound. However, the adiabatic \band{\Xstate}{\Cprimestate} bound feature is orders of magnitude weaker than the continuum bands presented here and an analysis on the change of character of bound-bound absorption bands with diabatisation is already provided above. The \band{\Xstate}{\Cprimestate} continuum bands for transitions to unbound \Cprimestate\ states above the $\rm S(^1D)+O(^3P)$ dissociation converge between both representations, since the non-adiabatic effects are far away from the peak at $\sim$\wav{78000} corresponding to vertical transitions from the electronic ground state. 
%For systems like \band{\Xstate}{\Cprimestate} we show the continuum structure to be completely unchanged by non-adiabatic effects since the region around the avoided crossing doesn't overlap well with the ground \Xstate\ and produces only a very shallow adiabatic potential well. 
%The major difference in this case is the different dissociation channels. 
However, if the avoided crossing occurred vertically above the \Xstate\ minima, we would expect non-adiabtaic effects to have a greater contribution to the continuum cross sections. 

%\red{DONT KNOW WHERE TO PUT THIS: These are the most extreme case of the impact of the non-adiabatic effects providing a strong metric to test the adequacy of non-adiabatic approaches.}

From the comparison above, we show that neglecting NACs within an adiabatic model can lead to drastic differences in the physics gleamed from the computed spectra.

%If one defined the NAC and DC properly, then these bands should become identical, however this plot shows the importance of treating NACs since their solutions are completely different and would describe completely different physics. 
%From our discussion in section \ref{subsec:DCs}, the diabatic representation of these bands are expected to be the more physical one. 

%Therefore, using the diabatic model one would be able to assign the \band{\astate}{\estate} band lines in the $E/hc \gtrsim$ \wav{50000} region.

%The effects to the spectrum caused by the diabatisation  can be understood through studying the PECs in Fig.~\ref{fig:ai_PECs} and coupling curves in Figs.~\ref{fig:ai_LSX}--\ref{fig:ai_DMY}. We have discussed that diabatisation removes steep gradients in couplings connecting the spatially degenerate states around the avoided crossing and  these couplings can lend intensities through mixing of the electronic wavefuctions. For example, we found that after diabatisation the vibrational structure of the \band{\Xstate}{\astate} between $0-10000$ \wav{} becomes more intense by $\sim 2$ orders of magnitude and its vibrational structure becomes more defined compared to calculations using the adiabatic model. This  may be due to some indirect coupling such as $\bra{\astate}{\rm SO}_z\ket{\Aprimestate}$ to which the \Aprimestate\ has spin-orbit couplings with both the \estate\ and \Cstate.

\section{Conclusions}

In this work, we use multireference methods of electronic structure theory combined with a diabatisation procedure to compute a fully diabatic model for the transient diatomic molecule sulphur monoxide. The model includes  23 spin-orbit, 23 (transition) dipole moment, and 14 electronic angular momentum curves for the \Xstate, \astate, \bstate, \Astate, \Bstate, \cstate, \Aprimeprimestate, \Aprimestate, \Cstate, \dstate, and \estate\ electronic states of SO and were produced \ai\ via CASSCF and MRCI methods using aug-cc-pV5Z basis sets. These curves were then used to compute the nuclear motion via solving the fully-coupled Schr\"{o}dinger equation with the \duo\ program. A further two electronic states (\Cprimestate\ and $(3)^1\Pi$) were computed along with their couplings, which are essential to forming the diabatic representation. 
%The on-the-fly diabatisation of our \ai\ model for SO removes the avoided crossings between the spatially degenerate pairs \estate--$(3)^1\Pi$ and \Cstate--\Cprimestate, consequently eliminating steep gradients within couplings to these states. 
%NACs are modelled using a geometric average between an initially refined Lorentzian and a Laplacian function using smoothness properties of the diabatic PECs. This averaged functional form of the NAC removes undesirable features of the Lorentizan and Laplacian, and models NACs very well.
The diabatisation procedure we present is a computationally low cost method to reconstruct the non-adiabatic couplings \emph{a priori} to nuclear motion calculations. 
To assess the importance of non-adiabatic effects for the spectroscopy of SO, we compare spectra computed in the diabatic and adiabatic representations without definition of NACs. The most notable difference  is the absence  of the  UV spectrum above $\sim$ \wav{50000} because of the illusionary predissociation from the adiabatic PECs. We also saw the adiabatically computed bound absorption bands to have lower intensities than the diabatic counterparts. It is therefore important to treat NACs for systems where these non-adiabatic interactions occur in spectroscopically important regions since they have drastic effects on spectral structure.

%We compute absolute intensities for every electro-rovibrational transitions between the lowest 11 diabatic singlet and triplet states of SO. 

%The coverage of our \ai\ model extends across the entire spectroscopic range where NACs are treated, providing a full description of the electronic structure of SO. Given the number of states and couplings we consider in this work, many couplings of which show non-adiabatic effects, we can produce non-LTE diabatic models which reproduce the structure of experimental spectra well. In all comparisons, we saw shifts in the line positions relative to experiment, which is to be expected at the MRCI level of theory (see section \ref{subsubsec:PEC}). \green{Considering the poor agreement between our MRCI computed \Xstate\ vibrational band centers to experiment, we compute the \Xstate\ PEC and dipole at a coupled-cluster level of theory which increased the line position accuracy 3 fold}. 

All coupling curves of SO are defined with self-consistent relative phases, which is crucial for spectral calculations \citep{14PaHiTe.AlO}. Therefore our spectroscopic model of SO provides a comprehensive and extensive theoretical baseline, which is the first fully reproducible spectroscopic description of SO longer than 147~nm.  Since the existing spectroscopic data on SO only covers \Xstate, \astate, \bstate, \Astate, and \Bstate, our \ai\ model can be used as  a benchmark for future rovibronic methods and calculations.

%Comparisons between the \band{\Cstate}{\Xstate}, \band{\estate}{\astate}, and \band{\Cprimestate}{\Xstate} spectral bands computed within the adiabatic and diabatic representations reveal diabatisation to have a great impact to assess the impact of diabatisation on the final computed spectra. For the former two bands, 

%In this paper, we not only show the importance of diabatisation in molecular dynamics calculations, but also provide a comprehensive diabatic model for the transient SO species which has no experimental coverage on electronic states energetically above \Bstate. Our model will provide strong groundwork for further refinement to experimental data, where we expect the accuracy of line positions and intensities to be improved, and also provide a strong baseline for predicting the SO spectrum at an \ai\ level up to $\sim$ \wav{60000}.

The topic of our next work will be to build a semi-empirical line list as part of the ExoMol project \citep{12TeYuxx.exo, 20TeYuAl} for SO through the refinement of our \ai\ model to experimental transition data, where we expect to reduce the shift in line positions relative to experiment. The final SO line list will have applications primarily in the atmospheric modelling of exoplanets \citep{98ZoFexx.SO,21HoRiSh, 12Krasnopolsky} and cool stars. Further applications of this empirical SO line list will be in shock zone modelling \citep{96Bachiller.ISM, 93ChMaxx.SO, 91AmElEl.SO}, SO lasing systems\citep{91MiYaSm.SO, 92StCaPo.SO}, and spectroscopy of Venus \citep{90NaEsSk.SO, 12BeMoBe.SO} and Io \citep{96Lellouch, 02deRoGr.SO}.

\section{Conflicts of interest}
There are no conflicts to declare.

\section*{Acknowledgements}

 This work was supported by UK STFC under grant ST/R000476/1. This work made  use of the STFC DiRAC HPC facility supported by BIS National E-infrastructure capital grant ST/J005673/1 and STFC grants ST/H008586/1 and ST/K00333X/1. We thank the European Research Council (ERC) under the European Union’s Horizon 2020  research and innovation programme through Advance Grant number  883830.

\section{Data availability}

The \ai\ curves of SO obtained in this study are phase consistent and are provided as part of the supplementary material to this paper along with our spectroscopic model in a form of a \duo\ input file. The line list computed with \duo\ can be directly used with the \exocross\ program to simulate the spectral properties of SO. We also provide a Julia source code used  for the diabatisation procedure is provided as part of the supplementary material.

\appendix
\section{Appendix}

\subsection{The Geometric average of DCs}

We show here the simple geometric average used to combine the DCs arising from our NACs modelled with a Lorentzian and a Laplacian. We have from Eq. (16) of \citet{15AnBaxx.diabat} the following
\begin{align*}
    V^{\rm ga}_{12} &= \sqrt{V^{\rm d,Lo}_{12}\cdot V^{\rm d,La}_{12}} \\
    &= \sqrt{\frac{1}{2}(V^{\rm a}_{2}-V^{\rm a}_{1}) \sin(2\beta^{\rm Lo})\cdot \frac{1}{2}(V^{\rm a}_{2}-V^{\rm a}_{1}) \sin(2\beta^{\rm La})} \\
    &= \frac{1}{2} (V^{\rm a}_{2}-V^{\rm a}_{1}) \sqrt{\sin(2\beta^{\rm Lo})\sin(2\beta^{\rm La})} \\
    &= \frac{1}{2} (V^{\rm a}_{2}-V^{\rm a}_{1}) \sin(2\beta^{\rm av})
\end{align*}   
where in the second line the matrix elements given in Eq.~\eqref{eq:dia_V} are substituted. Comparing the last two lines, one can recover Eq. \eqref{eq:beta_ga}, the mixing angle that corresponds to the geometrically averaged DC is then
\begin{align*}
\sin(2\beta^{\rm av}) = 
\sqrt{\sin(2\beta^{\rm Lo})\sin(2\beta^{\rm La})} \\
\rightarrow \beta^{\rm av} = \frac{1}{2} \arcsin\left(\sqrt{\sin(2\beta^{\rm Lo})\sin(2\beta^{\rm La})}\right) ,
\end{align*}  
which corresponds to Eq. (15) of \citet{15AnBaxx.diabat}.

\subsection{Diabatisation Procedure}
Non-adiabatic interactions cause steep gradients around the region of the avoided crossings in both PECs and coupling curves. To diabatise these couplings one requires the operation of the unitary transformation matrix given in Eq. \ref{eq:U(beta)}. For some coupling, $\zeta$, and considering two adiabatically interacting states $\ket{1}$ and $\ket{2}$, the exact transformation for non-diagonal couplings \brkt{1/2}{\zeta}{k}, where $k \neq 1,2$, from the adiabatic (`a') to the diabatic (`d') representation can be found as follows
\begin{gather*}
\begin{bmatrix} \brkteq{1}{\zeta}{k} \\ \brkteq{2}{\zeta}{k}\end{bmatrix}^{d} = \boldsymbol{U}^{\dagger}\begin{bmatrix} \brkteq{1}{\zeta}{k} \\ \brkteq{2}{\zeta}{k}\end{bmatrix}^{a}
\end{gather*}
However, diagonal couplings also require the off-diagonal counterparts, and transform similarly to the potential energies as,
\begin{multline*}
    \mathcal{D}^{\rm d} = \begin{bmatrix} \brkteq{1}{\zeta}{1}^{\rm d} & \brkteq{1}{\zeta}{2}^{\rm d} \\ \brkteq{2}{\zeta}{1}^{\rm d} & \brkteq{2}{\zeta}{2}^{\rm d} \end{bmatrix} \\
    = U^{\dagger}\begin{bmatrix} \brkteq{1}{\zeta}{1}^a & \brkteq{1}{\zeta}{2}^a \\ \brkteq{2}{\zeta}{1}^a & \brkteq{2}{\zeta}{2}^a \end{bmatrix} U = U^{\dagger}\mathcal{D}^{\rm a}U 
\end{multline*}
The computational procedure for performing the diabatisation is straightforward, and is based on the heuristic that the diabatisation should result in continuous potential energy curves that are twice differentiable due to the properties of the wave function and derivatives. For a function represented by a discrete grid of points $V_i$ at geometries $R_i$, gradients are calculated via finite differences and the task of satisfying this criterion is approximated by minimizing the sum of second derivatives of the function.

To achieve this we optimize the parameters of an arbitrary NAC function, $\phi(R; \{p\})$. The parameters $\{p\}$ are typically: a central geometry $R_{\rm c}$, and a characteristic width $\omega$ (see e.g Equations (\ref{eq:lorentzian}) and (\ref{eq:laplacian})). The NAC function, in turn, parameterises the transformation from the adiabatic potential energy operators $\boldsymbol{V}^\text{a}_i$ to the diabatic potential energy operators $\boldsymbol{V}^\text{d}_i$ at each geometry $i$, as described by Equation (\ref{eq:dia_V}).

The optimization itself is easy to achieve using any commonly available optimization libraries. In the present work we use the Julia programming language and the open-source Optim library's \texttt{Optim.minimizer} function to minimize the following loss function using the Nelder-Mead method \cite{Julia-2017, optim.jl_1.7.2}:
\begin{multline*}
     L\left(\omega, R_\text{c}; \{\boldsymbol{V}^\text{a}_i\}, \{R_i\}, \phi_\text{NAC}\right) \\ 
        = 2 \sum_{i = 2}^{N-1} \frac{1}
    {
    \left(R_{i+1} - R_{i}\right)\left(R_{i+1} - R_{i-1}\right)\left(R_{i} - R_{i-1}\right)
    } \\
    \times [\boldsymbol{V}^\text{d}_{i+1} \left(R_{i} - R_{i-1}\right) - \boldsymbol{V}^\text{d}_{i} \left(R_{i+1} - R_{i-1}\right) \\ + \boldsymbol{V}^\text{d}_{i-1} \left(R_{i+1} - R_{i}\right) ]
\end{multline*}
where the terms of the summation are simply central finite difference second derivatives of the diabatic potential energy operator.

The integration of the NAC function to obtain the mixing angle $\beta_i$ at each geometry is obtained by adaptive Gauss-Kronrod quadrature using the \texttt{QuadGK.jl} library.

Empirically we find that an initial guess for the characteristic width $\omega$ with the correct order of magnitude suffices for the optimizer to converge. However, the procedure is somewhat more sensitive to the initial guess for the central geometry $R_{\rm c}$ as a result we provide a convenience function that attempts to detect the central geometry by searching for the largest absolute value of the second order derivatives of the adiabatic operators. The source code for the diabatisation procedure is provided as part of the supplementary files.

\bibliographystyle{rsc}

\bibliography{./journals_phys,./atmos,./exoplanets,./ISM,./linelists,./methods,./programs,./SO,./SO2,./stars,./planets, ./partition, ./atomic, ./CS, ./Books, ./diatomic, ./abinitio, ./diabatisation,./sy, ./CO, ./software}

\end{document}